\documentclass[twocolumn]{aastex631}

\newcommand{\chioh}{\chi_\mathrm{oh}}
\newcommand{\chico}{\chi_\mathrm{co}}
\newcommand{\Pathchp}{Path$_\mathrm{CH^+}$}
\newcommand{\Pathoh}{Path$_\mathrm{OH}$}
\newcommand{\Pathwoh}{Path$_\mathrm{woH}$}

\newcommand{\taumid}{\tau_\mathrm{mid}}

\newcommand{\fracdg}{f_\mathrm{dg}}
\newcommand{\fracdgfid}{f_\mathrm{dg,fid}}

\newcommand{\NCtaumidfid}{ {\cal N}_\mathrm{C,\tau_\mathrm{mid}=1}^\mathrm{fid}}
\newcommand{\NCtaumid}{ {\cal N}_\mathrm{C,\tau_\mathrm{mid}=1}}

\newcommand{\nH}{n_\mathrm{H}}

\newcommand{\NH}{N_\mathrm{H}}
\newcommand{\nO}{n_\mathrm{O}}
\newcommand{\nC}{n_\mathrm{C}}
\newcommand{\NC}{N_\mathrm{C}}
\newcommand{\pcc}{\mathrm{cm}^{-3}}
\newcommand{\psc}{\mathrm{cm}^{-2}}
\newcommand{\Ac}{ {\cal A}_\mathrm{C}}
\newcommand{\Ao}{ {\cal A}_\mathrm{O}}

\newcommand{\nCp}{n(\mathrm{C^+})}

\newcommand{\nCO}{n(\mathrm{CO})}
\newcommand{\nCOana}{n(\mathrm{CO})_\mathrm{ana}}
\newcommand{\nCIana}{n(\mathrm{C^0})_\mathrm{ana}}
\newcommand{\NCO}{N(\mathrm{CO})}
\newcommand{\nCI}{n(\mathrm{C^0})}
\newcommand{\NCI}{N(\mathrm{C^0})}

\newcommand{\nHm}{n(\mathrm{H_2})}

\newcommand{\Qave}{\langle Q\rangle}

\newcommand{\FH}{F_\mathrm{H}}

\newcommand{\amin}{a_\mathrm{min}}
\newcommand{\amax}{a_\mathrm{max}}

\newcommand{\kohc}{k_\mathrm{oh,c}}

\newcommand{\aOH}{\alpha_\mathrm{oh}}
\newcommand{\koh}{k_\mathrm{o,h}}

\newcommand{\aCO}{\alpha_\mathrm{co}}

\begin{document}

\title{
A Constraint on the Amount of Hydrogen from the CO Chemistry in Debris Disks
}

\author[0000-0002-2707-7548]{Kazunari Iwasaki}
\affiliation{
Center for Computational Astrophysics, National Astronomical Observatory of Japan, Osawa, Mitaka, Tokyo 181-8588, Japan,
kazunari.iwasaki@nao.ac.jp
}

\author[0000-0001-8808-2132]{Hiroshi Kobayashi}
\affiliation{
Department of Physics, Nagoya University, Furo-cho, Chikusa-ku, Nagoya, Aichi 464-8602, Japan
}


\author[0000-0002-9221-2910]{Aya E. Higuchi}
\affiliation{
    Division of Science, School of Science and Engineering, Tokyo Denki University,
    Ishizaka, Hatoyama-machi, Hiki-gun, Saitama 350-0394, Japan
}

\author{Yuri Aikawa}
\affiliation{Department of Astronomy, Graduate School of Science, The University of Tokyo, 7-3-1 Hongo, Bunkyo-ku, Tokyo 113-0033, Japan}




\begin{abstract}

 The faint CO gases in debris disks are easily dissolved into C by UV
 irradiation, while CO can be reformed via reactions with hydrogen.
 The abundance ratio of C/CO could thus be a probe of the amount of hydrogen in the debris disks.
 We conduct radiative transfer calculations with chemical reactions for debris disks. 
 For a typical dust-to-gas mass ratio of debris disks, 
 CO formation proceeds without the involvement of H$_2$ 
 because a small amount of dust grains makes H$_2$ formation inefficient.
 We find that the CO to C number density ratio depends on a 
 combination of $n_\mathrm{H}Z^{0.4}\chi^{-1.1}$, where $n_\mathrm{H}$ is the hydrogen nucleus number density, 
 $Z$ is the metallicity, and $\chi$ is the FUV flux normalized by the Habing flux. 
 Using an analytic formula for the CO number density, we give constraints on the amount of hydrogen and metallicity for debris disks.
 CO formation is accelerated by excited H$_2$ either when the dust-to-gas mass ratio is increased or 
 the energy barrier of chemisorption of hydrogen on the dust surface is decreased.
 This acceleration of CO formation occurs only when the shielding effects of CO are insignificant. In shielded regions, the CO fractions are almost 
 independent of the parameters of dust grains.

\end{abstract}



\section{Introduction} \label{sec:intro}
Planets are believed to be formed in protoplanetary disks (PPDs) containing gas and dust.
The dissipation timescale of PPDs, especially that of gaseous component, is of
special importance for planetary system formation.
The gas giant planets need to be formed prior to the
significant dispersal of gas in disks \citep{Mizuno1978,Tanigawa2007}
and the gas dispersal may then lead to long term orbital instabilities of
protoplanets that trigger their impacts \citep{Iwasaki2001}.
The giant impacts between protoplanets are believed to form the terrestrial
planets in the solar system.

Infrared emission from PPDs decreases in several million
years \citep{Heisch2001}.
Older faint disks around main-sequence stars
are called debris disks \citep[][for review]{Hughes2018},
which might be explained by dust production due to collisional cascades starting from
disruptive collisions of kilometer or larger bodies.
That may be induced by planet formation events after
gas dispersal such as giant impacts for terrestrial planet formation
\citep[e.g.,][]{Genda2015} or planetary accretion in outer disks
\citep{Kobayashi2014}.
Sub-mm observations
showed some debris disks with ages of several 10\,Myrs still have
molecular CO gas \citep{Hughes2008,Dent2014,Moor2019}.
In the present work we aim to investigate if the gas in debris disks is a primary
origin, i.e., the gas of PPDs is not fully dispersed, or secondary origin.
This question is directly linked to how and when the disk gas dissipates.

In the secondary scenario, the observed CO is produced from solid bodies via collisions
\citep{Kral2016,Kral2017,Matra2015,Matra2017,Matra2018}.
However, CO is photo-dissociated rapidly.
For $\beta$
Pictoris whose disk has CO with $\sim 3 \times 10^{-5} M_\oplus$, the
dissociation timescale is about $\sim 100$ years so that the mass
production of CO during the stellar age is estimated to be $\sim 3
M_\oplus$, where $M_\oplus$ is the Earth mass \citep[cf.][]{Kral2016}.  The IR
observations for comets by AKARI infer the fraction of CO to H$_2$O
$\sim 0.1$ \citep{Ootsubo2012}.
    Total solid bodies required for the observed gas
      is simply estimated from the CO mass and fraction to be $30 M_\oplus$,
    which is comparable to or larger than that of the solar system.

The gas production due to sublimation strongly depends on the
temperature of parent bodies.
Using the condensation temperature, the sublimation timescale
is estimated to be 0.003~yrs for kilometer-sized CO bodies with 50~K,
while it may be somehow longer for mixed ices. However,
the timescale is much shorter than the ages of debris disks.
It would thus be difficult to keep $\sim 3M_\oplus$ of CO in the icy bodies, 
unless there was a steady inflow from the outer colder regions.
Carbon dioxide, CO$_2$, can be another reservoir.
CO observed around comets in our solar system is mainly explained by the photo-dissociation of
CO$_2$ \citep{Weaver2011} so that
CO might be less abundant than CO$_2$ in comets.
  Outgassing CO$_2$ changes into CO via photo-dissociation.
Collisional vaporization occurs due to shock
heating \citep{Ahrens1972,Kurosawa2010}, so that high speed impacts
($\la 5$\,km/s) are required even for vaporization of volatile materials
\citep[e.g.,][]{Okeefe1982,Kraus2011}.
The collisional velocities are estimated to be
$~1~\mathrm{km~s^{-1}}(e/0.1) (M_*/M_\odot)^{1/2} (r/10~\mathrm{au})^{1/2}$,
where $e$ and $r$ are the orbital eccentricities and semimajor axes of colliding bodies, respectively,
$M_*$ is the mass of the central star, and $M_\odot$ is the solar mass
(e.g., Kobayashi \& L\"ohne 2014).
The collisional velocities of bodies with $e\sim 0.1$ beyond 10\,au are much slower than the vaporization velocity.
The collisional outgassing of CO$_2$ from cometary bodies is therefore difficult beyond 10 au,
where CO gas is observed in debris disks.
In summary, in the secondary-origin scenario, a large amount of reservoir ice is required,
while it should be sublimated efficiently beyond $>10~$au.

If the gas in debris disks is a primary origin, or is a mixture of primary and secondary
components, CO can be reformed from C$^+$ and C atoms in the gas phase.
The required CO reservoir is then significantly reduced compared with the case of
purely secondary gas. \citet{Higuchi2017} showed that the abundance ratio of atomic carbon to
CO can be a probe of the abundance of H$_2$ relative to carbon, i.e. the metallicity of gas in debris disks.
If the gas is a remnant of PPDs , the metallicity is expected to be close
to that of the interstellar medium (ISM).
On the other hand, if gas is supplied from solid bodies, the metallicity is
much larger than the ISM value while hydrogen is provided by H$_2$O desorbed from solid bodies.

Although the CO chemistry in debris disks was investigated by many authors
\citep{Kamp2000,Kamp2003,Gorti2004,Hughes2008,Roberge2013}, the elemental abundances in their models are not different
from those in the ISM significantly.
The metallicity can be different from the ISM value, depending on the origin of gas.
Thus, in this paper, we examine how the CO chemistry depends on metallicity
by calculating detailed chemical reactions and thermal processes,
and the radiative transfer including exact line transfers in the debris disks.

This paper is organized as follows:
one-dimensional plane-parallel PDR calculations.
The results of the PDR calculations are described 
and an analytical formula for the amount of CO is developed on the basis of 
the plane-parallel PDR calculations in Section \ref{sec:result}.
Astrophysical implications are discussed in Section \ref{sec:discuss}.
Finally, our results are summarized in Section \ref{sec:summary}.

\section{Methods and Models}\label{sec:model}

\subsection{Numerical Methods}\label{sec:method}
The Meudon PDR code version 1.5.4 ({\ttfamily http://pdr.obspm.fr/}) is a publicly
available photon dominated region (PDR) code which is designed to model
a stationary plane-parallel slab of gas and dust controlled by
far-ultraviolet (FUV) photons \citep{LeP2006}.
A plane-parallel slab is illuminated by a radiation field penetrating from
the left-side of the slab.  

The Meudon code solves the radiative
transfer at each point in the slab, taking into account dust extinction
and line absorption of gaseous species.
We use the exact method described in \citet{Goi2007} 
which allows us to take into account overlapping of the H, H$_2$, and CO UV absorption lines.
For the H$_2$ line transfer, we define a value of $J_\mathrm{max}=3$
under which the exact method is used.
The FGK approximation proposed by \citet{Fed1979}
is used for all electric transitions for $J_l\ge J_\mathrm{max}$, where $J_l$ is 
the lower level of Lyman and Werner transitions.
We do not take into account the exact line transfer for $^{13}$CO and C$^{18}$O.

Coupled with the radiative transfer, 
the full level population system of a number of species, including H$_2$ and CO, is computed 
from the detailed balance between 
radiative transitions, collisional transitions, transitions due to the cosmic microwave background, 
formation and destruction processes 
both in the gas phase and on dust grains \citep{LeP2006}.

This treatment allows us to evaluate the H$_2$ self-shielding effect more accurately than
using the fitting formula provided by \citet{DraineBertoldi1996}, which 
has been used in most studies.
Furthermore, the accurate calculation of the level populations provides 
the accurate rates of chemical reactions associated with excited H$_2$, 
which will be discussed in Section \ref{sec:parametersurvey}.
Similarly, the shielding effect of CO is calculated more accurately 
than using the table from \citet{Visser2009}.

The Meudon code calculates the thermal equilibrium of gas and dust grains.
For gas temperature, heating processes (including the
photo-electric heating on dust, exothermic chemical reactions, and cosmic
rays) are balanced with cooling processes (including infrared and
millimeter emission from ions, atoms, and molecules).
The treatment of dust grains is explained in
Section \ref{sec:dust}.
Chemical and thermal equilibria determine the abundances of each chemical species at
each position.

In order to investigate 
the basic properties of the chemical 
structure of debris disks, 
in Section \ref{sec:result}, we conduct 
simple plane-parallel PDR calculations 
of a semi-infinite gas slab 
with a uniform density illuminated by stellar radiation fields and 
the interstellar radiation field (ISRF).

We should note that the Meudon PDR code is not 
applicable directly to the multi-dimensional 
disk geometry because 
it is designed for the one-dimensional plane-parallel geometry.
In debris disks, we should take into account the geometrical dilution of the 
stellar radiation and vertical ISRF.
Nevertheless, we will find that the main 
underlying physics of debris disks 
can be approximated by the findings from the one-dimensional plane-parallel PDR calculations.
In Section \ref{sec:observation}, 
we compare the observational results with the predictions obtained 
by applying the findings of the plane-parallel models to disk models.

\subsection{Modifications to the Meudon Code}\label{sec:mod}
We made some modifications to the Meudon code because 
it is not designed for debris disks.


Firstly, we modified the calculation of the photo-dissociation of CO$^+$.
In calculating the photo-dissociation rate of species, the Meudon code 
divides the species into two groups.
For one group, 
the photo-dissociation rate is estimated from 
a fitting formula, which is a form of $\propto \exp(-\beta A_\mathrm{V})$, 
assuming the radiation field 
to be the ISRF \citep{vanDishoeck2006}, where $\beta$ is the fitting parameter.
The rate is scaled by the ratio of the total UV flux to the ISRF flux.
While this formulation is efficient and is often used in PDR calculations of debris disks,
it can be inaccurate when the SED of the radiation field is different from the ISRF.
In addition, the possible shielding of photo-cross sections that resonate with absorption lines 
is not considered in the exponential form.
Since dust grains in debris disks are larger than those found in the ISM,
$\beta$ is expected to be inaccurate.

For the other group, the photo-dissociation rate is computed 
directly from $\int \sigma_\lambda I_\lambda/h\nu d\lambda$;
$\sigma_\lambda$ is the photo-dissociation cross section and $I_\lambda$ is the radiation intensity 
summed over all incidence angles. 
This treatment allows us to evaluate the photo-dissociation 
consistent with the local radiation field at each position.
It is however computationally expensive and used only for important species.

In the Meudon PDR code, CO$^+$
was included in the former group where the 
fitting formula is used
although CO$^+$ is one of the most important 
species to form CO in debris disks.
In this work, the photo-dissociation rate of CO$^+$ is calculated from the numerical integration 
of $\int \sigma_\lambda I_\lambda/h\nu d\lambda$,
where $\sigma_\lambda$ is taken from 
\citet{Lavendy1993} \citep[also see][]{Heays2017}, and tabulated as a function of the wavelength.

One of the important molecules to form CO in our results is the hydroxyl radical OH, whose 
photo-dissociation rate is computed directly from the integration of photo-reaction cross sections.
It is dissociated by photons with wavelengths longer than 1100~\AA.
In the adopted radiation fields, the photo-dissociation rate of OH is 
mainly determined by the $1 ^2\Sigma^-$ channel, which 
absorbs photons with 
$\sim 1600~$\AA \citep{vanDishoeck1983}.
We will define a normalized UV flux dissociating OH in Section \ref{sec:fuv}.

Secondly, we modified the rate coefficient of an endothermic reaction 
of $\mathrm{O+H_2\rightarrow OH + H}$, which is one of the most important 
reactions to form CO.
The Meudon PDR code treats endothermic reactions of C$^{+}$, S$^+$, O, and OH with H$_2$,
taking into account excited states of H$_2$.
Except for reactions of 
$\mathrm{C^++H_2}$ and $\mathrm{S^++H_2}$ \footnote{
To estimate the enthothermic chemical reaction rates,
the Meudon PDR code uses
the results of the theoretical studies done by 
\citet{Zanchet2013} and \citet{Herrez-Aguilar2014} 
for $\mathrm{C^++H_2\rightarrow CH^+ + H}$
and by \citet{Zanchet2013_Sp} 
for $\mathrm{S^++H_2\rightarrow SH^+ + H}$.
}
the rate coefficients were calculated 
assuming that all the internal energies are used to overcome 
an activation barrier or an endothermicity.
In order to calculate the reaction rate of $\mathrm{O+H_2\rightarrow OH + H}$ more accurately,
we use the fitting formulae for the rate coefficient of 
$\mathrm{O+H_2}(v)\rightarrow \mathrm{OH} + \mathrm{H}$ for $v=0$, 1, 2, and 3 
derived by \citet{Agndez2010} based on the theoretical calculations by 
\citet{Sultanov2005}. 
The reaction rates for $v>3$ are replaced by that for $v=3$, 
where $v$ is the vibrational quantum number.
As a result, the total reaction rate may be underestimated.
Recently, precise rates especially for $v>3$ 
have been estimated by \citet{Veselinova2021}\footnote{
O and H$^*_2$ with $v>3$ does not affect the CO formation because 
the H$^*_2~(v>3)$ fractions are too small to affect the CO formation.
}.

Thirdly, we do not consider heating owing to the di-electronic recombination.
In the Meudon code, for simplicity, the electron captured in an upper electronic level
is assumed to be de-exited only through collisions, resulting in gas heating.
The density range considered in this paper is so low that 
most of transitions in the cascade of the captured electron will occur through 
spontaneous emissions. 
We only consider gas cooling owing to the removal of the kinetic energy of recombined electron.

\subsection{Incident Radiation Fields}\label{sec:fuv}

We consider two kinds of radiation fields which are incident on the edge 
of a semi-infinite gas slab.

One is the ISRF 
\citep{Habing1968,Draine1978,Mat1983}.
The ISRF is given by the summation of four components.
One is the ISRF from far- to near-ultraviolet;
an expression of the first component is given by
\begin{eqnarray}
I_\mathrm{ISRF}(\lambda) &=& 107.192 \left( \frac{\lambda}{1~\mathrm{\AA}}\right)^{-2.89}\nonumber\\
&& \times \left[
\mathrm{tanh}\left( 4.07\times10^{-3}\left( \frac{\lambda}{1~\mathrm{\AA}} \right)
- 4.5991 \right)+1  \right] \nonumber\\
&& \mathrm{erg~cm^{-2}~s^{-1}~\AA^{-1}~sr^{-1}}
\label{ISRF}
\end{eqnarray}
for $\lambda \ge 912~$\AA
\citep{Mat1983}\footnote{Equation (\ref{ISRF}) is a fitting function of the results of \citep{Mat1983} 
as shown in the manual of the Meudon PDR code.}.
The second ISRF component is supplied by cold stars and
is expressed as a combination of three black bodies at 6184, 6123, and 2539~K.
The third component comes from the dust thermal emission which is estimated 
by the DustEM code \citep{Compigne2011}.
The forth ISRF component is the cosmic microwave background radiation given 
by a black body at 2.73~K.

The other radiation field is the stellar emission taken from
ATLAS9 \citep{Kurucz1992,Howarth2011}.
We consider two spectral types of A5V and A1V.
The effective temperatures $T_\mathrm{eff}$ of the A5V and A1V stars are
8250~K and 9000~K, respectively.
The metallicities of the stars are set to the solar value.
The surface gravity is fixed to $\log_{10}g = 4$.
Examples of the A5V and A1V stars are $\beta$~Pictoris 
\citep{Lecavelier2001}
and 49~Ceti \citep{Roberge2013}, respectively.

\begin{figure}[htpb]
        \centering
        \includegraphics[width=8cm]{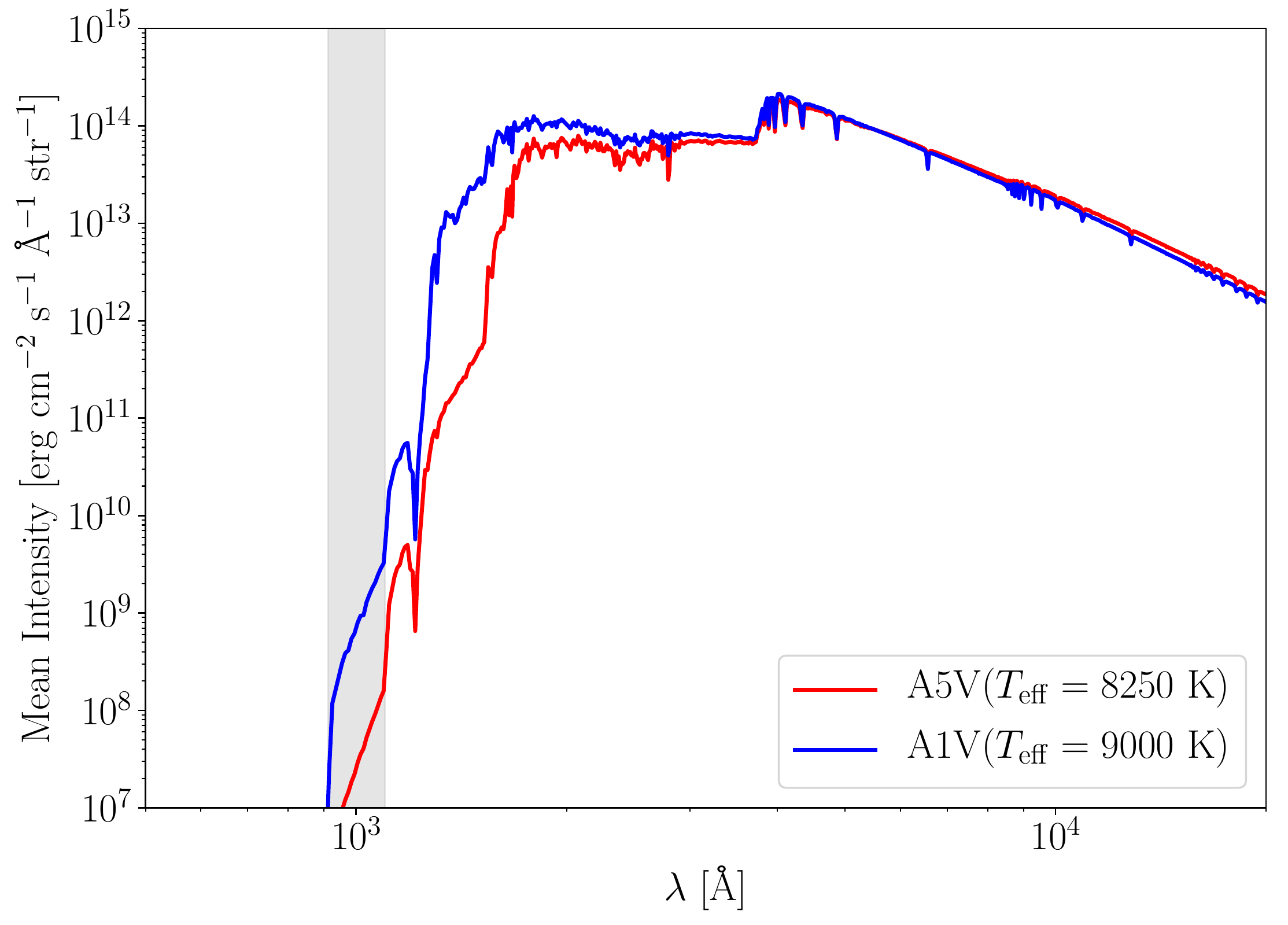}
\caption{
Mean intensities of the stellar radiation of A5V (red) and A1V (blue) stars 
at a representative distance of 50~au from the central star 
as a function of wavelength.
  The FUV photons within the gray region induces photo-dissociation of H$_2$ and CO,
  and photo-ionization of C$^0$.
}
\label{fig:spec}
\end{figure}

As the stellar radiation incident into the slab, 
the radiation field at a representative 
distance of 50~au from the star is adopted.
The mean intensities at $50~$au are shown in  Figure \ref{fig:spec}.



We parameterize the UV fluxes that dissociate species which are important to form CO.
The following two ranges of wavelengths are focused on.

One wavelength range is $912~\mathrm{\AA}<\lambda<1100~\mathrm{\AA}$, which is shown by
the gray region in Figure \ref{fig:spec}.
The photons in this wavelength range photo-dissociate CO and H$_2$, and photo-ionize C$^0$.
The FUV photon number flux integrated over the wavelength range
is denoted by $F_\mathrm{FUV,CO}$, and is often measured in units of the \citet{Habing1968} field
flux, $\FH\equiv 1.2\times 10^7$~cm$^{-2}$~s$^{-1}$ 
\citep{Bertoldi1996}.
The normalized incident FUV photon number 
flux $\chico \equiv F_\mathrm{FUV,CO}/F_\mathrm{H}$\footnote{
In this paper, $\chi_\mathrm{co}=1$ refers to an FUV field whose 
energy density at the surface of the gas slab is equal to that 
of the Habing field considering all $4\pi$ steradian in the free-space.
}
is given by
\begin{equation}
        \chico = \chi_\mathrm{co,star} + \chi_\mathrm{co,ISRF}/2,
      \label{chi}
\end{equation}
where 
$\chi_\mathrm{co,star}$ indicates the normalized stellar FUV flux at a distance of 
50~au from the central star, and 
$\chi_\mathrm{co,ISRF}=1.3$ is the normalized ISRF flux derived from Equation (\ref{ISRF}).  

{
The other wavelength range is considered to 
characterize the OH photo-dissociation rate.
The cross section of photo-dissociation of OH adopted in the Meudon PDR 
code has the two broad peaks centered around $\lambda\sim 1090~\mathrm{\AA}$ ($3~^2\Pi$)
and $\lambda\sim 1600~\mathrm{\AA}$ ($1~^2\Sigma^-$) \citep{vanDishoeck1983,vanDishoeck1984,Heays2017}.
The latter peak mainly contributes to the OH photo-dissociation while the contribution from the former peak
is negligible. This is because 
as shown in 
Figure \ref{fig:spec} the intensities of the stellar radiation increase rapidly 
with wavelength in the corresponding wavelength range.
Considering that the intensities of the stellar radiation  
are an increasing function of wavelength within the 
latter peak,
we set a wavelength range of
$1600~\mathrm{\AA}\le \lambda \le 1700~\mathrm{\AA}$,
which corresponds to the 
long wavelength half of the latter peak.
The OH photo-dissociation 
rate is roughly proportional to the FUV photon number flux 
integrated over this wavelength range, which 
is denoted by $F_\mathrm{FUV,OH}$.
}
For convenience, we use the normalized FUV flux $\chioh \equiv F_\mathrm{FUV,OH}/10^{12}~\mathrm{cm^{-2}~s^{-1}}$.

From the stellar spectra of the A5V and A1V stars shown in 
Figure \ref{fig:spec}, one derives that 
$\chico$ are $1.4$ and $31$, respectively.
The contribution of the ISRF to $\chioh$ is negligible, and 
$\chioh=2.4$ for A5V and $24$ for A1V, respectively. 

In this paper, the models with ($\chico=1.4$, $\chioh=2.4$) and 
those with ($\chico=31$ and $\chioh=24$) are referred to 
"weak-FUV" and "strong-FUV", respectively.



\subsection{Dust Properties}\label{sec:dust}
Dust grains play important roles in at least three processes.
First, they are responsible for absorption and scattering of photons.
Second, they work as a catalyst in some chemical reactions.
For instance, H$_2$ forms on the surface of dust grains rather than by gas phase reactions.
Third, they contribute to thermal processes through photo-electric heating and
collision with the gas.
These processes depend significantly on the dust properties:
{the size distribution, dust-to-gas mass ratio, and chemical compositions.}

\subsubsection{Size Distribution}\label{sec:size}

One of the differences of the debris disk from the ISM is
absence of small dust grains.
The small dust grains with radii smaller than $\sim 1~\mu$m are blown out by
the radiation pressure of stellar photons
\citep{Burns1979,Kobayashi2008,Kobayashi2009}.
Absence of the small dust grains reduces the visual extinction,
the H$_2$ formation rate, and the photo-electric heating rate 
for a given dust-to-gas mass ratio.
For simplicity, we consider spherical dust grains whose radii are denoted by $a$.
The power-law grain size distribution $n(a)\propto a^{-3.5}$ is assumed, where
$n(a)da$ is the number density of the dust grains
with the radius range from $a$ to $a+da$ \citep{Mathis1977}.
This power-law size distribution is controlled by
the collisional cascade \citep{Dohnanyi1969,Tanaka1996},
although the power-law index could be modulated due to the size dependence of collisional strength
\citep{Kobayashi2010}.
In this work, the minimum and maximum radii of the dust grains are fixed to
$a_\mathrm{min}=1~\mu$m and $a_\mathrm{max}=10~\mu$m.
We will discuss that our results do not depend 
directly on the size distribution in Section \ref{sec:plane}.

\subsubsection{Composition}

The dust grains are assumed to be a mixture of graphite and silicon.
The absorption and scattering coefficients of dust grains 
are calculated by averaging those of the graphite and silicon 
in a ratio of $7:3$, 
taking into account the size distribution.
The heat capacity of the dust grains, 
which is used to determine the dust temperature at each 
dust size and position, is computed 
assuming the mixed composition.
The absorption/scattering coefficients and heat capacity at each dust size
are computed by \citet{LaorDraine1993} and \citet{DraineLi2001}.

\subsubsection{Charge of Dust Grains}\label{sec:charge}
In addition to the FUV flux, 
the charge of dust grains characterizes the photo-electric yield on 
the dust grain.
The Meudon code takes into account the probability distribution function 
of the dust charge for each dust size $f(a,Q)$, which is determined by 
collisional charging and photo-electric ionization \citep{BakesTielens1994}, where
$Q$ is the dust charge.

As the grain size increases, the photo-electric ionization rate increases faster than 
the recombination rate. 
As a result, the charge of dust grains increases with $a$.
Since the maximum grain size considered in this work is up to 1~cm,
a great number of the charge grid is required.

To reduce the computational cost,
we approximate the charge distribution function at each grain size $f(a,Q)$ 
by a delta function of $\delta(Q-\Qave_a)$, where $Q$ is the 
dust charge and $\Qave_a$ is the mean charge for the grain size $a$.
The validity of the approximation $f(a,Q)=\delta(Q-\Qave_a)$ is confirmed.

\subsubsection{Dust-to-gas Mass Ratio}\label{sec:fracdg}
  Since the dust-to-gas mass ratio $\fracdg$ of debris disks is unknown observationally and
  theoretically, a wide range of $\fracdg$ is considered in the present work.
      Using the size distribution $n(a)$, 
  the dust-to-gas mass ratio is expressed as 
  \begin{equation}
      \fracdg \equiv \frac{1}{\mu_\mathrm{C}m_\mathrm{C}\nC} 
      \frac{4\pi}{3}\rho_\mathrm{gr}\langle a^3\rangle n_\mathrm{d},
      \label{fdg0}
  \end{equation}
  where 
  \begin{equation}
      \langle g(a) \rangle \equiv \frac{1}{n_\mathrm{d}} 
      \int_{\amin}^{\amax}g(a) {n}(a)da,
  \end{equation}
  $n_\mathrm{d} = \int_{\amin}^{\amax} {n}(a)da$ is the total number density of dust grains,
  $m_\mathrm{x}$ and $n_\mathrm{x}$ are the mass and number density of 
  element ``x'', respectively,
  $\mu_\mathrm{x}$ is the mean molecular weight per element ``x'' nucleus, and 
  $\rho_\mathrm{gr}=2.62~\mathrm{g~\pcc}$ is the internal density of the dust grains. 
  We should note that the gas mass density is expressed as $\mu_\mathrm{C}m_\mathrm{C}\nC$
  because the gas component is considered on the basis of carbon.
  The gas metallicity is denoted by $Z$, and 
  the elemental abundances of the gas phase used in the present work are shown in Table \ref{tab:abundance} 
  for the solar metallicity $Z=1$.
  When $Z$ changes, the relative abundances among metals are fixed.   
  From Table \ref{tab:abundance}, $\mu_\mathrm{C}$ is given by 
  \begin{equation}
          \mu_\mathrm{C}(Z)=  883\left( Z^{-1} + 6.6\times 10^{-3}\right).
      \label{muC}
  \end{equation}

\begin{table}[htpb]
	\centering
	\begin{tabular}{|c|c|c|}
		\hline
        atom & relative abundance 
        of the gas phase & reference \\
		\hline
		\hline
    He & $0.1$ & \\
		\hline
    C & {$1.32\times 10^{-4}$} & 1 \\
		\hline
    N & {$7.5\times 10^{-5}$}& 2 \\
		\hline
    O & {$3.2\times 10^{-4}$}& 3 \\
		\hline
    Ne & {$6.9\times 10^{-5}$} & 4 \\
		\hline
    Si & {$8.2\times 10^{-7}$}& 5 \\
		\hline
    S & {$1.0\times10^{-8}$}& 6 \\
		\hline
    Ar & {$3.29\times 10^{-6}$}& 7 \\
		\hline
    Fe & {$1.5\times 10^{-8}$} & 1 \\
		\hline
	\end{tabular}
	\caption{
        Relative elemental abundances  of the gas phase with respect to hydrogen for $Z=1$.
    (1) \citet{Savage1996}, (2) \citet{Meyer1997}, (3) \citet{Meyer1998},
      (4) \citet{Adamkovics2011}, (5) \citet{Morton1975}, (6) \citet{Tieftrunk1994},
      (7) \citet{Lodders2008}.
  }
	\label{tab:abundance}
\end{table}

  In the context of PPDs, $n_\mathrm{d}$ is often determined from Equation (\ref{fdg0}) at a given $\fracdg$.
  By contrast, we use the fact that in debris disks, the amount of dust grains is constrained by 
  the optical depth at {\it V} band,
  which is often estimated from the fraction of the disk luminosity to stellar luminosity.
  The most practical criterion for debris disks is 
  $\tau < 8\times 10^{-3}$ \citep{Hughes2018}, where $\tau$ denotes a typical vertical optical depth of debris disks.
  A typical optical depth along the mid-plane 
  $\taumid$ is given by $\tau \theta^{-1}$, where 
  $\theta$ is the aspect ratio of debris disks.

  The mean free path $\ell$ of a photon for dust absorption 
  depends on the grain size distribution as follows:
  \begin{equation}
      \ell = \left(n_\mathrm{d} \langle Q_\mathrm{abs} \pi a^2\rangle\right)^{-1}. 
      \label{meanfreepath}
  \end{equation} 
  In the integration, the absorption coefficient $Q_\mathrm{abs}$ 
  can be set to be unity because
  $2\pi a/\lambda > 1$ is satisfied for $a\ge 1~\mu$m at the {\it V} band.
  Eliminating $n_\mathrm{d}$ using Equations (\ref{fdg0}) and (\ref{meanfreepath}), 
  one obtains 
  \begin{equation}
  \fracdg = 1.74\times 10^{23}~\mathrm{cm}^{-3}~\frac{\langle a^3\rangle}{\langle a^2\rangle}
  \mu_\mathrm{C}^{-1}  {\cal N}_\mathrm{C,\taumid=1}^{-1},
    \label{fdg}
  \end{equation}
  where ${\cal N}_\mathrm{C,\taumid=1} = \nC \ell$ 
  is the mid-plane carbon nucleus column density  of the gas phase
  at $\taumid=1$ of the dust opacity at {\it V} band.
  The values of $\fracdg$ adopted in our PDR calculations will be shown in
  Section \ref{sec:modelparameter}.

    As mentioned in Section \ref{sec:size}, 
    the size distribution $n(a)\propto a^{-3.5}$ is used, 
    and $\langle a^3\rangle /\langle a^2\rangle = \sqrt{\amin\amax}$.
    
\subsubsection{Temperature of Dust Grains}\label{sec:Tgr}

In the Meudon PDR code, the temperature of dust grains $T_\mathrm{gr}$ is 
computed from the energy balance between absorption of the radiation and 
thermal emission at each grain radius bin.
For reference, we here show the grain temperature given by 
\citet{Hughes2008} as follows:
\begin{equation}
    T_\mathrm{gr} = 10^2~\mathrm{K}\left( \frac{L_*}{10L_\odot} \right)^{0.2} 
    \left( \frac{a}{1~\mu\mathrm{m}}\right)^{-0.2} 
    \left( \frac{r}{50~\mathrm{au}} \right)^{-0.4},
    \label{Tgr}
\end{equation}
where $L_*$ is the stellar luminosity, 
$a$ is the grain radius, and 
$r$ is the distance from the central star.
The dust temperatures obtained the PDR calculations 
are  consistent with Equation (\ref{Tgr}).

\subsubsection{Grain Surface Chemistry}\label{sec:dustsurface}

The Meudon code takes into account H$_2$ and HD formation with the Langmuir-Hinshelwood (LH) 
and the Eley-Rideal (ER) mechanisms \citep{LeBourlot2012}.
The grain-surface chemistry for other species heavier than H$_2$ and HD 
is not included in this study because photo-desorption is so efficient that heavy species are difficult to freeze out on the grain surface 
\citep[e.g.,][]{Grigorieva2007}.

In debris disks, the most important H$_2$ formation mechanism on dust grains 
is the ER mechanism involved by chemisorption of H atoms.
This is because the dust temperatures, which are as high as 
$\sim$100~K as shown in Equation (\ref{Tgr}),
are so high that a hydrogen atom physisorbed on the grain surface 
evaporates before it combines with another hydrogen atom.

In the ER mechanism, there are several free parameters 
in the Meudon PDR code.
One is a sticking coefficient $\alpha(T)$, which approaches zero for high temperatures 
because the hydrogen atom cannot stick on the dust grains owing to the excess kinetic energy.
The Meudon PDR code adopts the empirical form 
$\alpha(T)=1/(1+(T/T_\mathrm{stick})^\beta)$, 
where $T_\mathrm{stick}=464~$K and $\beta=3/2$ are adopted \citep{LeBourlot2012}.

Another parameter is $T_\mathrm{chem}$ corresponding to the 
energy barrier that a hydrogen atom overcomes 
to reach a chemisorption site on the grain surface.
The H$_2$ formation rate is highly sensitive to $T_\mathrm{chem}$
because the probability of overcoming the energy 
barrier is proportional to $\exp(-T_\mathrm{chem}/T)$.
Nonetheless, $T_\mathrm{chem}$ is highly uncertain.  
Its value depends on the surface condition and composition of dust grains.
The energy barrier of chemisorption onto a perfect graphite surface 
is as high as $\sim 2000$~K
\citep[e.g.,][by using quantum mechanics]{Sha2002}.
This has been confirmed by 
the experimental study of a highly ordered pyrolytic graphite 
in the laboratory done by \citet{Zecho2002}.
By contrast, topological defects on the carbonaceous surface 
decrease $T_\mathrm{chem}$, and can make chemisorption 
almost barrierless 
($T_\mathrm{chem}\sim 10-100~$K), 
depending on structures \citep{Ivanovskaya2010}.
Experiments of chemisorption of H on porous, defective, 
aliphatic carbon surfaces have 
found that the activation energy is 
as low as $\sim 70$~K \citep{Mennella2006}. 
By contrast, in the cases with silicate, 
quantum chemical calculations revealed 
$T_\mathrm{chem}\sim 300~$K both for crystalline silicate 
\citep{Navarro-Ruiz2014} and amorphous silicate \citep{Navaroo-Ruiz2015},
but they have not been confirmed experimentally yet.

Uncertainty of the H$_2$ formation rate on 
warm dust grains also has been 
discussed in PDRs illuminated by strong radiation 
from massive stars.
Except for the absence of small dust grains, 
the environments are similar to debris disks.
In such PDRs, \citet{Habart2011} observationally found that 
rotationally-excited H$_2$ is more abundant than the predictions of PDR models, 
suggesting that the H$_2$ production rate on warm dust grains 
needs to be higher than a typical value \citep[e.g,][]{Jura1974}.
This result suggests that there is significant uncertainty in H$_2$ formation rates 
on warm dust grains.

As will be mentioned in Section \ref{sec:modelparameter}, 
considering the uncertainty of $T_\mathrm{chem}$, 
$T_\mathrm{chem}$ is changed as one of the 
model parameters in our PDR calculations.

\subsection{Model Parameters}\label{sec:modelparameter}

  We summarize the model parameters in our models and show their parameter spaces.
  The model parameters are divided into ones related to the radiation field,
  the dust, gas components, the H$_2$ formation on the grain surface.
  The model parameters are tabulated in Table \ref{tab:param}.

\subsubsection{The Parameters Related to the Radiation Fields, $(\chi_\mathrm{co},\chi_\mathrm{oh})$}

As mentioned in Section \ref{sec:fuv}, 
in order to investigate how the CO chemistry depends on radiation fields,
the weak-FUV and strong-FUV models are considered.
They are characterized by $\chico$ and $\chioh$.




\subsubsection{The Parameter Related to the Amount of Dust Grains, $f_\mathrm{dg}$}

   The amount of dust grains is characterized by $\NCtaumidfid$.
   We here estimate  $\NCtaumidfid$ from observation results of $\beta$ Pictoris and 49 Ceti, which 
  are examples of the stars having gas-poor and gas-rich debris disks, respectively.
      Since both the debris disks are almost edge-on, the observed column densities are comparable to 
      those along the mid-plane.
      The total column density of species ``A'' integrated along the mid-plane is denoted by ${\cal N}(A)$.
  
  For $\beta$ Pictoris, several authors derive the C$^0$ column densities from the [C{\sc I}] $^3P_1$-$^3P_0$ emission line
  \citep{Higuchi2017,Cataldi2018} and $^3P$ absorption lines \citep{Roberge2000}.
  Their values range from ${\cal N}(\mathrm{C}^0)\sim 2\times 10^{16}$ to $7\times 10^{16}~\psc$.
  The C$^+$ column densities are estimated to be $\sim 2\times 10^{16}-12\times 10^{16}~\psc$ from the [C{\sc I$\!$I}] 
  158~$\mu$m emission line \citep{Cataldi2014,Cataldi2018} and 
  absorption lines \citep{Roberge2006}.
  The CO column densities are estimated to be $3\times 10^{14}-5
  \times 10^{14}~\psc$ from emission lines \citep{Higuchi2017} 
  and $\sim 6\pm 0.3\times 10^{14}~\psc$ from
  absorption lines \citep{Roberge2000}.
  These observations suggest that ${\cal N}(\mathrm{C}^0)$ is comparable to ${\cal N}(\mathrm{C}^+)$, and 
   ${\cal N}(\mathrm{CO})$ is roughly two orders of magnitude smaller than ${\cal N}(\mathrm{C}^0)$ and  ${\cal N}(\mathrm{C}^+)$.
  The carbon nucleus column density ${\cal N}_\mathrm{C}$ is estimated to be $\sim {\cal N}(\mathrm{C}^0) + {\cal N}(\mathrm{C}^+)\sim 10^{17}~\psc$. 
  The spatial distribution of the dust surface area obtained by HST observations of \citet{Heap2000} is fitted by \citet{Fernandez2006}.
  The mid-plane optical depth is about $\taumid \sim 10^{-2}$, 
  where the fitting formula is integrated over $50~\mathrm{au}\le R\le 120~\mathrm{au}$, which 
  is the gas extent of the best fit model in
  \citet{Cataldi2018}.
  Finally, $\NCtaumid^\mathrm{\beta Pic}\sim {\cal N}_\mathrm{C}/\taumid \sim 10^{19}~\psc$ is obtained for $\beta$ Pictoris.

  For 49 Ceti, the C$^0$ and CO column densities are estimated to 
  be ${\cal N}(\mathrm{C}^0)\sim 
  2\times 10^{18}~\psc$ \citep{Higuchi2019} and 
  ${\cal N}(\mathrm{CO})
  \sim 1.8\times 10^{17}-5.9 \times 10^{17}~\psc$ \citep{Higuchi2020}, 
  respectively. 
  The C$^+$ column density is not well constrained because only a lower limit of the C$^+$ mass, 
  which is smaller than the CO mass, was obtained \citep{Roberge2013}.
  We however expect that  ${\cal N}_\mathrm{C}$ 
  is comparable to ${\cal N}(\mathrm{C}^0)$ because 
  ${\cal N}(\mathrm{C}^+)$
  should not be much larger than 
   ${\cal N}(\mathrm{C}^0)$ if all CO are formed 
  through chemical reactions.
  The vertical 
  optical depth estimated from the fractional luminosity is 
  $\tau \sim  10^{-3}$ \citep{Jura1993,Jura1998}.
  Assuming that the aspect ratio of the dust distribution is $\sim 0.1$, the mid-plane optical depth is $\taumid \sim 10^{-2}$.
  Finally, we obtain $\NCtaumid^\mathrm{49Ceti}\sim 2\times 10^{20}~\psc$ for 49 Ceti.

  From $\NCtaumid^\mathrm{\beta Pic}=10^{19}~\psc$
  and $\NCtaumid^\mathrm{49Ceti}=2\times 10^{20}~\psc$, 
  a fiducial value of $\NCtaumid={\cal N}_\mathrm{C}/\tau_\mathrm{mid}$ is set to $ \NCtaumidfid=2\times 10^{19}~\psc$
  ($\NCtaumid^\mathrm{\beta Pic}=0.5\NCtaumidfid$, $\NCtaumid^\mathrm{49Ceti}=10\NCtaumidfid$) 
  because the behavior of the CO formation changes around 
  ${\cal N}_\mathrm{C,\taumid=1}\sim \NCtaumidfid$ 
  for weak-FUV and strong-FUV 
  as will be shown in Section \ref{sec:result}.



In this paper, instead of $\NCtaumid$, a normalized dust-to-gas mass ratio is used as the 
parameter indicating the amount of dust grains. Using Equation (\ref{fdg}), one obtains 
\begin{equation}
\frac{\fracdg}{\fracdgfid} =  \left( \frac{\NCtaumid}{\NCtaumidfid} \right)^{-1},
    \label{fdg_tau_Nc}
\end{equation}
where $\fracdgfid$ is the dust-to-gas mass ratio at  $\NCtaumid=\NCtaumidfid$ and 
given by 
  \begin{eqnarray}
    \fracdgfid(\amin,\amax,Z) &=& 3\times 10^{-3}
    \left\{ Z^{-1} + 6.6\times 10^{-3} \right\}^{-1} \nonumber \\
 &&   \times \left( \frac{\amin}{1~\mathrm{\mu m}} \right)^{1/2}
    \left( \frac{\amax}{10~\mathrm{\mu m}} \right)^{1/2},
    \label{fdgfid}
  \end{eqnarray}
where $n(a)\propto a^{-3.5}$ is used.
We should note that although both $\fracdg$ and $\fracdgfid$ depend on $\amin$ and $\amax$, 
the ratio $\fracdg/\fracdgfid$ is independent of the size distribution for a given  $\NCtaumid$.

To study how the amount of dust grains affects the thermal and chemical 
structures, 
the PDR calculations are performed with three different  $\fracdg$ as 
listed in Table \ref{tab:param}.
For weak-FUV, since a typical $\fracdg$ is about $\sim 2\fracdgfid$
for $\beta$ Pictoris, the cases with 
($0.1\fracdgfid$, $\fracdgfid$, $10\fracdgfid$) are considered.
For strong-FUV, since a typical $\fracdg$ is about $\sim 0.1\fracdgfid$, 
the cases with ($10^{-2}\fracdgfid$, $0.1\fracdgfid$, $\fracdgfid$)  
are considered.


\subsubsection{The Parameters Related to the Gas Component, ($\nC, Z$)}

The parameters associated with the gas component are $\nC$ and metallicity $Z$.
Given $\nC$ and $Z$, the corresponding hydrogen nucleus number density is 
 $\nH = \nC (\Ac Z)^{-1}$,
where $\Ac=1.32\times 10^{-4}$ is the carbon elemental abundance of the gas for $Z=1$ 
(Table \ref{tab:abundance}).
The hydrogen gas density is difficult to constrain 
in most debris disks observationally
although there are some successful exceptions  which will be described in Section \ref{sec:caveat}.

Fiducial carbon nucleus number densities are defined as $\nC=1.32\times 10^2~\pcc$ for weak-FUV 
and $\nC=1.32\times 10^3~\pcc$ for strong-FUV.
The debris disk around $\beta$ Pictoris has a spatial extent of $\sim 70~\mathrm{au}$ in the best-fit model in 
\citet[][also see Section \ref{sec:betaPic}]{Cataldi2018}.
A mean density is $\nC\sim 10^{17}~\psc/70~\mathrm{au}\sim 100~\pcc$, which is comparable to the fiducial $\nC$ for weak-FUV.
For 49 Ceti, because the radial extent is $\sim 100~$au, 
which corresponds to a cut-off radius of the surface density distribution 
\citep[][also see Section \ref{sec:49Ceti}]{Hughes2018},
the mean density is $\nC \sim 2\times 10^{18}~\psc/100~\mathrm{au} 
\sim 10^3~\pcc$, which is comparable to 
the fiducial $\nC$ for strong-FUV.
For comparison, the densities 10 times larger than the typical $\nC$ are also 
considered for both the weak-FUV and strong-FUV models.


Next, we consider a range of the gas metallicity.
The gas metallicity in debris disks depends on its origin. 
If the gas is a remnant of PPDs, the gas metallicity is 
expected not to be significantly different from the solar metallicity $Z=1$. 
It is maximized when all the gas comes from the secondary processes; hydrogen is provided by photo-dissociation of H$_2$O outgassed from 
dust grains. In this case, the number of hydrogen nuclei becomes comparable to that of oxygen nuclei, 
and the gas metallicity reaches $1.6\times 10^{3}$ since ${\cal A}_\mathrm{O} = 3.2\times 10^{-4}$ at $Z=1$ (Table \ref{tab:abundance}).
In this paper, a range of $1\le Z\le 10^3$ is considered. 

\subsubsection{The Parameter Related to the H$_2$ Formation on the Grain Surface, $T_\mathrm{chem}$}

In order to investigate how the uncertainty of the H$_2$ formation affects CO formation (Section \ref{sec:dustsurface}),  
the cases with $T_\mathrm{chem}=300$~K and $10$~K are considered.
The fiducial case $T_\mathrm{chem}=300~$K leads 
to a moderate H$_2$ formation rate, corresponding to situations where 
chemisorption efficiently occurs on grains with plenty of surface defects \citep{LeBourlot2012}. 
The cases with $T_\mathrm{chem}=10~$K give an upper limit of the H$_2$ formation rate.
Other parameters $T_\mathrm{stick}$ and $\beta$ are fixed to $464$~K and 1.5, respectively, because
the gas temperatures do not exceed $\sim 500$~K in most cases.

\begin{table*}[htbp]
    \begin{center}
    \begin{tabular}{|l|c|c|}
        \hline
        parameter  & weak-FUV & strong-FUV \\
        \hline
        \hline
        FUV flux (CO, H$_2$, C$^+$) $\chico$ & $1.4$ & $31$ \\
        \hline
         FUV flux (OH) $\chioh$ & 
        $2.4$ & $24$  \\
        \hline
        dust-to-gas mass ratio $\fracdg$ & 
        $0.1\fracdgfid$, $\fracdgfid$, $10\fracdgfid$ &
        $0.01\fracdgfid$, $0.1\fracdgfid$, $\fracdgfid$  
        \\
        \hline
        gas density $\nC~[\pcc]$ 
        & 
        $1.32\times 10^2~\pcc, 1.32\times 10^3~\pcc$ 
        &
        $1.32\times 10^3~\pcc, 1.32\times 10^4~\pcc$ \\
        \hline
        the gas metallicity $Z$ & \multicolumn{2}{c|}{$1,10,10^2,10^3$} \\
        \hline
        {
        \begin{minipage}{5cm}
        $T_\mathrm{chem}$~[K]$^{(1)}$
        \end{minipage}
        } & 
         \multicolumn{2}{c|}{ 300, 10} \\
        \hline
    \end{tabular}
    \caption{ 
   {  $^{(1)}$The energy barrier of  chemisorption.}
    List of the model parameters considered in the plane-parallel PDR calculations}.
    \label{tab:param}
    \end{center}
\end{table*}

\section{Results}\label{sec:result}

\begin{figure*}[htpb]
        \centering
       \includegraphics[width=18cm]{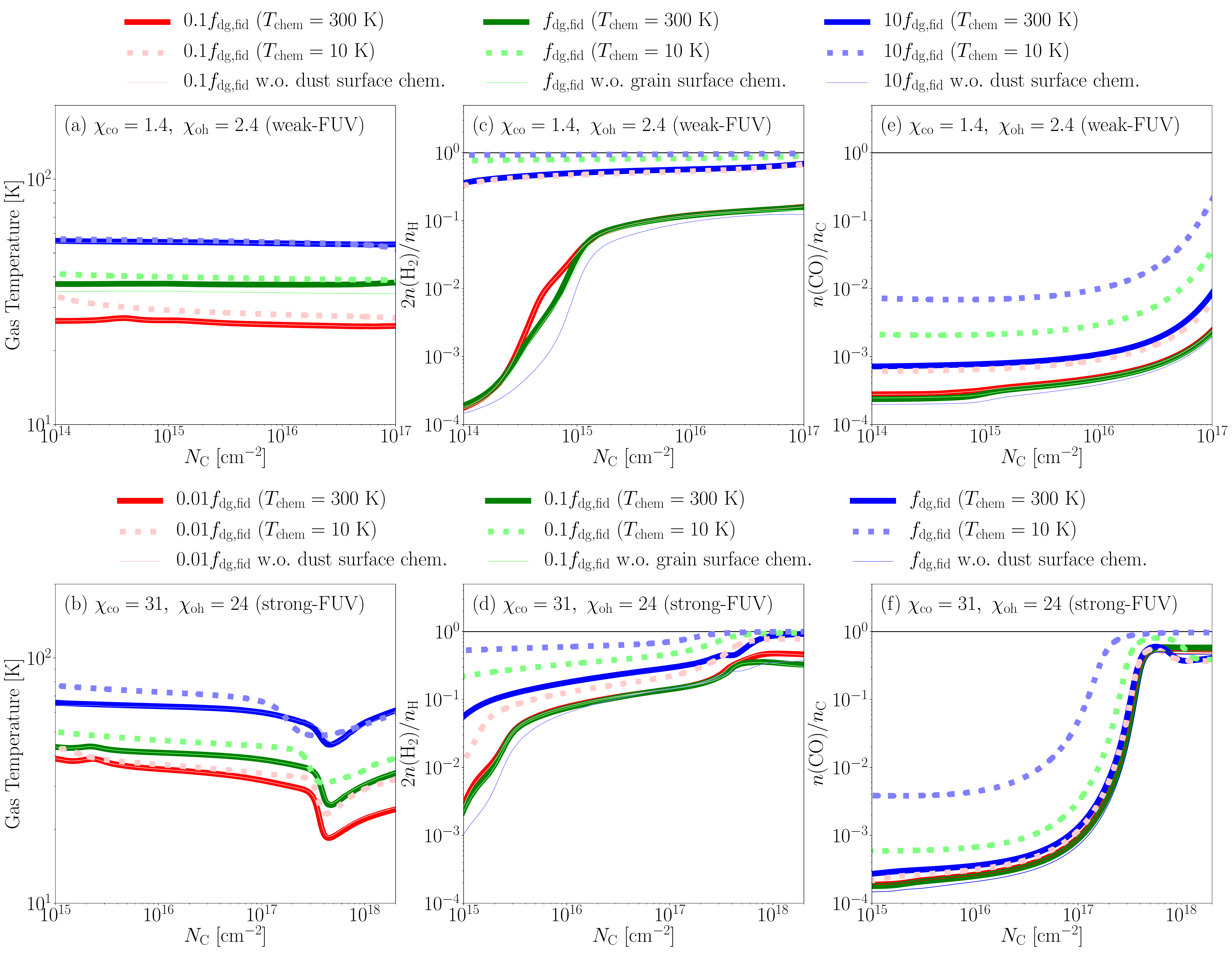}
\caption{
    Results of the PDR calculations where the incident flux 
    corresponds to the summation of ISRF and 
    stellar radiation 
    flux at a representative distance of 50~au from the central star.
    Profiles of the gas temperature 
    (left column), $2\nHm/\nH$ (middle column), and  $\nCO/\nC$ (right column) as 
    a function of $\NC$.
        The top and bottom panels show the results for weak-FUV and strong-FUV models, respectively.
        The gas metallicity is fixed to $Z=1$, and 
        the gas densities are fixed to $\nC=1.32\times 10^2~\pcc$ for weak-FUV and $\nC=1.32\times 10^3~\pcc$ for strong-FUV. 
        The colors represent $\fracdg$ which are $(0.1\fracdgfid,\fracdgfid, 10\fracdgfid)$ for weak-FUV and
        $(0.01\fracdgfid,0.1\fracdgfid,\fracdgfid)$ for strong-FUV. 
        The thick solid and dotted lines correspond to the results with $T_\mathrm{chem}=300~$K and $10~$K, respectively.
        The thin solid line indicates the results without the H$_2$ formation on  the grain surface. 
}
   \label{fig:plane}
\end{figure*}

\subsection{Overall Behaviors of the Models with the Solar Metallicity
}\label{sec:plane}

In this section, we investigate how the results of
the fiducial models 
($\nC=1.32\times 10^2~\pcc$ for weak-FUV and 
$\nC=1.32\times 10^3~\pcc$ for strong-FUV)
depend on the dust parameters $(\fracdg/\fracdgfid,  T_\mathrm{chem})$ 
and incident FUV fluxes ($\chico$, $\chioh$).
The gas metallicity is set to $Z=1$.

We should note that 
the results with different $\amin$ and 
$\amax$ are not presented
because our results are independent of 
both $\amin$ and $\amax$ as long as 
$\fracdg/\fracdgfid$ and $\nC$ 
remain unchanged.
This is explained as follows.
The grain-surface chemistry and photo-electric heating 
from dust grains  are characterized by 
the total geometric cross section of dust grains $n_\mathrm{d} \langle \pi a^2\rangle$.
  Equations (\ref{meanfreepath}) and (\ref{fdg_tau_Nc}) give 
  \begin{equation}
      n_\mathrm{d}\langle \pi a^2\rangle
      = \frac{n_\mathrm{C}}{ \NCtaumid}
      = \left( \frac{\fracdg}{\fracdgfid} \right) \frac{\nC}{ \NCtaumidfid},
      \label{crosssec}
  \end{equation}
  where we use the fact that $Q_\mathrm{abs}=1$ because $Q_\mathrm{abs}\sim 1$ for $a > 1~\mu$m.
  Equation (\ref{crosssec}) shows that $n_\mathrm{d}\langle \pi a^2\rangle$ 
  is controlled by $\nC$ and $\fracdg/\fracdgfid$. 
  Since $\NCtaumid^\mathrm{fid}$
  is a given quantity, 
  $n_\mathrm{d}\langle \pi a^2\rangle$ 
  does not depend on the size distribution of 
  dust grains 
  if $\nC$ and $\fracdg/\fracdgfid$ are fixed.
Thus, the thermal and chemical processes related to dust grains, i.e., 
H$_2$ formation on dust grains and photo-electric heating,
are independent of the size distribution for fixed $\nC$ and $\fracdg/\fracdgfid$.

The results are shown in Figure \ref{fig:plane}.
In this paper, instead of the optical depth, $\NC$ is used 
as a measure of the spatial coordinate,
where $\NC$ is the carbon nucleus column density integrated from the slab edge to a given position.
In Figure \ref{fig:plane}, the maximum column densities are set 
to $\NC=10^{17}~\psc$ for weak-FUV and $\NC=2\times 10^{18}~\psc$ for strong-FUV 
according to the observational constraints shown in Section \ref{sec:modelparameter}.
\subsubsection{Gas temperature}\label{sec:planeT}
Figures \ref{fig:plane}a and \ref{fig:plane}b show 
that the gas temperatures increase with $\fracdg$ for 
both the weak-FUV and strong-FUV models
because the photo-electric heating rate from dust grains, which 
is one of the dominant heating processes, increases in proportion to $\fracdg/\fracdgfid$.
A weak positive dependence of $T$ on $T_\mathrm{chem}$ 
is attributed to enhancement of the heating rate owing to the 
H$_2$ formation on the grain surface.

Comparison between Figures \ref{fig:plane}a and \ref{fig:plane}b shows that 
the gas temperatures increase with $\chico$ because 
the photo-electric heating rate increases with $\chico$.

\subsubsection{Molecular Hydrogen}\label{sec:planeH2}
Before presenting the results, we introduce the analytic formation rate 
of H$_2$ on grain surfaces, 
\begin{equation}
    F_\mathrm{H_2,dust} = 3\times 10^{-3}R_\mathrm{H_2,ISM}~Z
    \left( \frac{\fracdg}{\fracdgfid} \right) \sqrt{T} \kappa(T) \nH n(\mathrm{H})
  \label{RER}
\end{equation}
\citep{LeBourlot2012},
where Equation (\ref{crosssec}) is used, $R_\mathrm{H_2,ISM}=3\times 10^{-17}~$cm$^{3}$~s$^{-1}$ 
is a typical rate coefficient in the ISM \citep{Jura1974}, and 
$\kappa(T)$ is the chemisorption efficiency\footnote{
We should note that the Meudon code does not solve Equation (\ref{RER}) directly, but 
solves a set of rate equations considering chemisorbed hydrogen atoms and the dust size distribution
\citep[see][for details]{LeBourlot2012}.
In addition to the chemisorbed H atoms, the Meudon code treats H$_2$ formation with physisorbed H atoms.
}.
\citet{LeBourlot2012} derived 
\begin{equation}
        \kappa(T) = \frac{\alpha(T)\exp\left( -T_\mathrm{chem}/T \right)}
        {1+\alpha(T)\exp\left( -T_\mathrm{chem}/T \right)},
    \label{kappa}
\end{equation}
where $\alpha(T)$ and $T_\mathrm{chem}$ are defined in Section \ref{sec:dustsurface}.
For gas temperatures lower than $T_\mathrm{stick}=464~$K, which is 
satisfied in all the cases shown in Figure \ref{fig:plane}, 
the H$_2$ formation rate is proportional to $\exp(-T_\mathrm{chem}/T)$, and 
it depends sensitively on $T$.

First, the cases with $T_\mathrm{chem}=300~$K are considered.
Although H$_2$ mainly forms on grain surfaces for the ISM, this may not be the case 
for debris disks because the H$_2$ formation rate is extremely small.
In order to examine the contribution of the grain-surface chemistry to H$_2$ formation, 
we performed the additional PDR calculations where the grain-surface chemistry 
is switched off and plotted the results by the thin lines in 
the middle column of Figure \ref{fig:plane}.
For $\fracdg\le\fracdgfid$ (weak-FUV) 
and $\fracdg\le 0.1\fracdgfid$ (strong-FUV), 
there are almost no changes in the H$_2$ fractions with and without
the grain-surface chemistry (Figures \ref{fig:plane}c and \ref{fig:plane}d), 
indicating that the grain-surface chemistry does not contribute to H$_2$ formation.
In this case, H$_2$ is mainly formed in the gas phase,
and the dominant chemical reaction is 
$\mathrm{CH^+ + H\rightarrow H_2 + C^+}$,
where CH$^+$ is formed by the radiative association between C$^+$ and H\footnote{
Especially for metal-poor environments, the so-called H$^-$ channel is an important H$_2$ formation path \citep[e.g.,][]{Sternberg2021}.
Additional PDR calculations were performed with chemical reactions associated with H$^-$, which 
had not been activated in the default setting of the Meudon code.
The photo-detachment of H$^-$ is computed directly from $\int \sigma_\lambda I_\lambda/h\nu d\lambda$ (Section \ref{sec:mod}),
where the cross section obtained in \citet{Wishart1979} is used while the Meudon code calculates
it by using a fitting formula.
We confirmed that the H$^-$ channel does not affect the CO chemistry.
}.

The H$_2$ formation rate of the grain-surface chemistry 
is enhanced in proportion to $\fracdg$ (Equation (\ref{RER})) while
the rate in the gas phase does not change.
As shown in Figures \ref{fig:plane}c and \ref{fig:plane}d, 
$\fracdg$ is high enough for H$_2$ to be formed mainly through 
the grain-surface chemistry when $\fracdg=10\fracdgfid$ for weak-FUV and 
$\fracdg=\fracdgfid$ for strong-FUV.

For $T_\mathrm{chem}=300~$K, 
the H$_2$ fractions stay low level even after 
they increase rapidly around $\sim 10^{14}~\psc$ owing to self-shielding \citep{DraineBertoldi1996}. 
This is because the H$_2$ formation rate is so low that the destruction 
reactions in the gas phase and H$_2$ photo-dissociation keep the H$_2$ fractions low.




As mentioned in Section \ref{sec:dustsurface}, there 
is large uncertainty in H$_2$ formation on the warm dust grains with 
$T_\mathrm{gr}\sim 100~$K.
A decrease in $T_\mathrm{chem}$ from 300~K to 10~K 
significantly enhances the H$_2$ fractions
in Figures \ref{fig:plane}c and \ref{fig:plane}d 
because $\kappa(T)$ depends sensitively on $T_\mathrm{chem}$.
Considering the gas temperatures shown in Figures \ref{fig:plane}a and \ref{fig:plane}b, 
the H$_2$ formation rate increases by a factor of 
$\sim 1.6\times 10^4$ at $T=30$~K and $\sim 1.3\times 10^2$ at $T=60$~K.

\subsubsection{Carbon Monoxide, CO}\label{sec:planeCO}


First, the models with $T_\mathrm{chem}=300~$K are considered.
Figures \ref{fig:plane}e and \ref{fig:plane}f show that 
$\nCO/\nC$ does not depend on $\fracdg$ significantly 
as long as 
$\fracdg\le \fracdgfid$ for both weak-FUV and strong-FUV. 
The CO fractions decrease with increasing $\chico$ because 
stronger FUV flux destroys CO more efficiently.

When $\fracdg$ is increased from $\fracdgfid$ to $10\fracdgfid$, 
the weak-FUV models show that 
the CO fractions increase by a factor of 2 or 3.

Acceleration of the CO formation owing to H$_2$ is more significant 
for $T_\mathrm{chem}=10~$K than for $T_\mathrm{chem}=300$~K since 
a decrease in $T_\mathrm{chem}$ increases the H$_2$ fraction (Section \ref{sec:planeH2}).
The CO fractions monotonically increase with $\fracdg$. 

The $\fracdg$-dependence of the CO fractions disappears 
when the grain-surface chemistry is omitted from the PDR calculations.
This clearly indicates that 
the enhancement of the CO fraction for larger $\fracdg$ is related to 
efficient H$_2$ formation on the grain surface.

How much H$_2$ is needed to accelerate CO formation?
Comparing between the middle and right columns of Figure \ref{fig:plane},
one can see that H$_2$ fraction must be well above 0.1 for the CO fractions 
to increase by more than on order of magnitude.
When the H$_2$ fraction is comparable to 0.1 as in the strong-FUV models 
with ($\fracdg=0.01\fracdgfid$, $T_\mathrm{chem}=10~$K)
and ($\fracdg=\fracdgfid$, $T_\mathrm{chem}=300~$K), 
CO formation is not accelerated significantly.

In the deep interior $\NC>5\times 10^{16}~\psc$, 
the CO fractions increase owing to shielding effects, 
which will be discussed in Section \ref{sec:shielding}.

\subsection{CO Formation in H$_2$-poor Environments}\label{sec:analyticmodel}

In Section \ref{sec:result}, we show that the CO formation is independent of the amount of H$_2$ 
when H$_2$ formation is inefficient in 
situations where $T_\mathrm{chem}$ is high and/or 
$\fracdg$ is small.
In this section, we 
investigate how CO forms in such 
H$_2$-poor environment, and 
develop an analytic formula for the CO fraction that is 
applicable in a wide range of $\NC$ 
in Sections \ref{sec:COana} and \ref{sec:shielding}.


\subsubsection{Dependence of CO Fractions on Gas Metallicity}\label{sec:COZ}

\begin{figure*}[htpb]
\centering
\includegraphics[width=15.0cm]{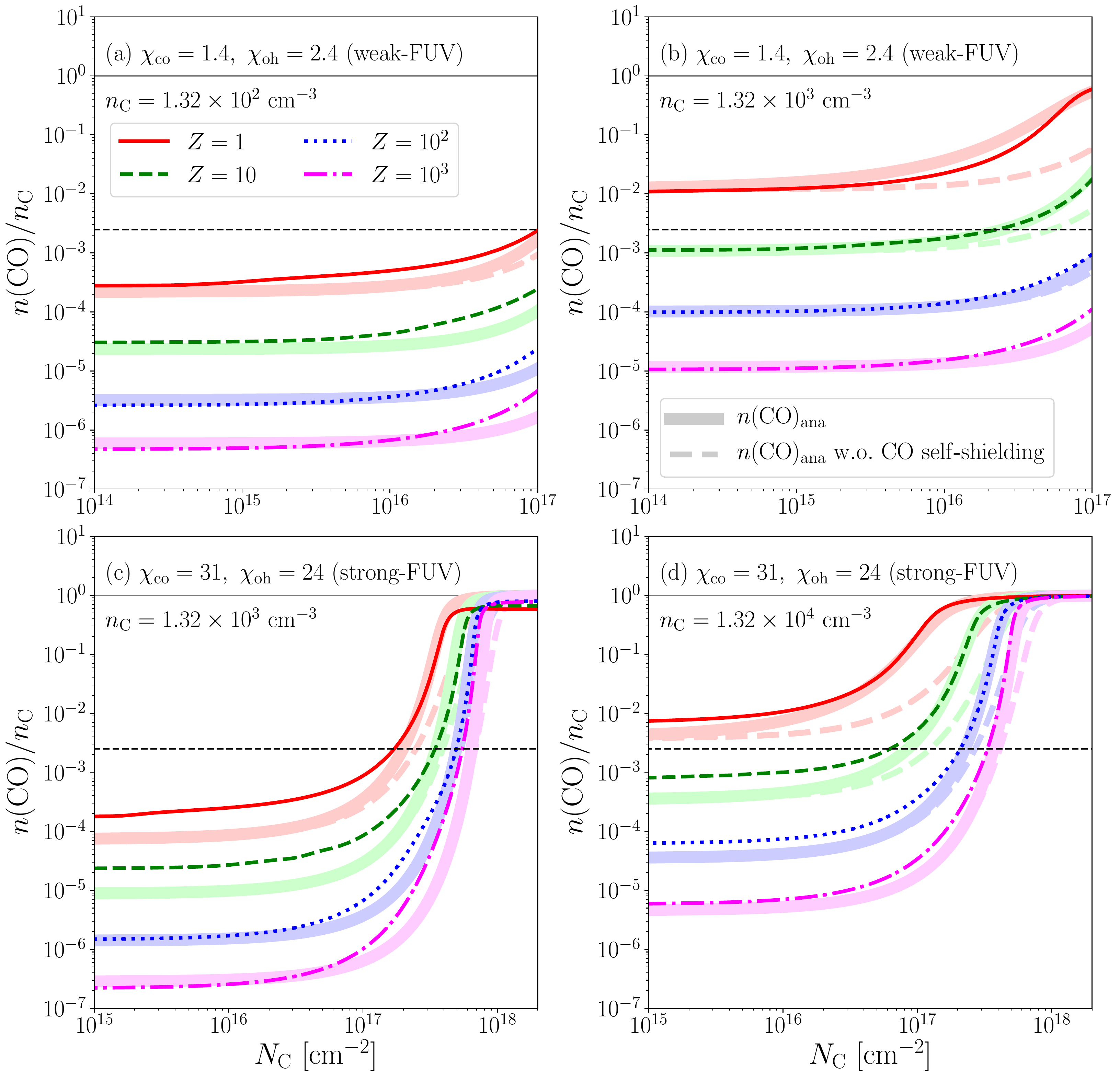}
\caption{
            The CO fractions as a function of $N_\mathrm{C}$ for 
            (a) weak-FUV ($\nC=1.32\times 10^2~\pcc$), 
            (b) weak-FUV ($\nC=1.32\times 10^3~\pcc$),
            (c) strong-FUV ($\nC=1.32\times 10^3~\pcc$), and  
            (d) strong-FUV ($\nC=1.32\times 10^4~\pcc$).
            The difference in the gas metallicites is shown by the red solid ($Z=1$), green dashed
            ($Z=10$), blue dotted ($Z=10^2$), and 
            magenta dot-dashed ($Z=10^3$) lines.
           The horizontal black dashed lines indicate
           $(\nCO/\nC)_\mathrm{cri}$ (Equation (\ref{COshield_cri})).
           The thick lines with lighter colors show $\nCOana$ as 
           a function of $N_\mathrm{C}$
           taking into account both the C$^0$ and CO shielding effects.
           The dashed lines with lighter colors 
           correspond to $\nCOana$ 
           without the CO self-shielding effect.
}
\label{fig:COshield}
\end{figure*}

Figure \ref{fig:COshield} shows the spatial distributions 
of the CO fractions for various $\chi$, $\nC$, and $Z$ listed in Table \ref{tab:param}.
The dust-to-gas mass ratios are fixed to 
$\fracdg=0.1\fracdgfid$ for weak-FUV and 
$\fracdg=0.01\fracdgfid$ for strong-FUV, where 
the CO formation proceeds, regardless of H$_2$ (Figure \ref{fig:plane}).

We first focus on the CO fractions for $\NC\lesssim 10^{17}~\psc$.
For given $\nC$ and the FUV fluxes ($\chico,\chioh$), 
the CO fractions decrease with increasing $Z$.
Since $Z$ is changed keeping $\nC$ constant, $\nH$ decreases with increasing $Z$. 
The CO fractions thus decrease with decreasing $\nH$.
This behavior has been pointed out by \citet{Higuchi2017}; 
hydrogen is involved in CO formation. 

Comparing between the panels of Figure \ref{fig:COshield}, 
one can see that the CO fractions increase as $\nC$ increases and/or $\chico$ decreases.
This behavior was qualitatively expected because the CO formation rate will
increase with increasing $\nC$ and the CO destruction rate will decrease with decreasing $\chico$.

For all the models shown in Figure \ref{fig:COshield}, 
the CO fractions begin to increase in deep interiors
owing to shielding effects, 
which will be investigated in Section \ref{sec:shielding}.

\subsubsection{An Analytical Formula for the CO Fraction}\label{sec:COana}

\begin{figure}[htpb]
        \centering
        \includegraphics[width=9cm]{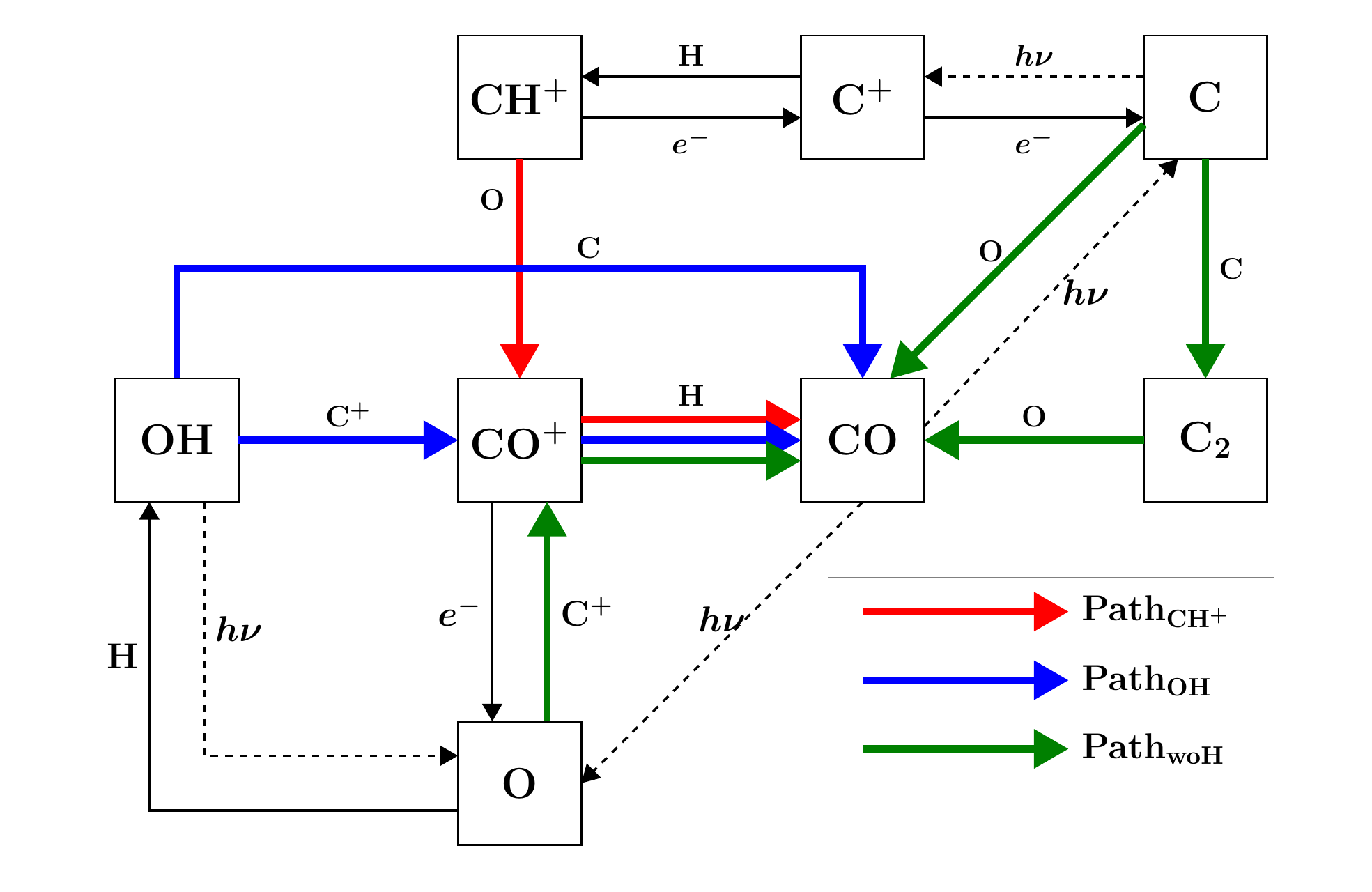}
\caption{
     The most important reactions to form CO in the H$_2$-free gas.
     The red and blue arrows show the CO formation paths intermediated by CH$^+$ and OH, respectively.
     The green arrows correspond to the CO formation paths not associated with hydrogen.
     The photo-dissociation of OH and CO and photo-ionization of C are shown by the dashed arrows.
     The chemical reaction rate coefficients are summarized in Table \ref{tab:reaction}.
}
\label{fig:COnetwork}
\end{figure}

In this section, we develop an analytic formula 
for the CO fractions from
the most important chemical reactions associated with CO formation 
of the H$_2$-free gas in Figure \ref{fig:COnetwork}.
There are three main paths to form CO.
At the high density limit (Equation (\ref{nHreq}))
, the CO formation is intermediated mainly by OH  ({\Pathoh}). 
For lower densities, the dominant CO formation path depends on $Z$.
At $Z\le 10$, the CO formation is intermediated by CH$^+$ (\Pathchp).
For higher $Z$, the CO formation proceeds without hydrogen as shown by the green arrows
in Figure \ref{fig:COnetwork} (\Pathwoh).


The analytic formula taken into account 
all the chemical reactions shown in Figure \ref{fig:COnetwork} is complicated and is not useful.
In Appendix \ref{app:COana}, we construct a fitting formula of the CO fraction 
based on {\Pathoh}, which is dominated for higher densities,
as follows:
\begin{equation}
    \frac{\nCOana}{\nC} = \frac{\Ao}{ {\cal A}_\mathrm{O,ism}} 
    \left(
        10^{-14} \eta^{1.8} + 6.0\times 10^{-11} \eta 
    \right),
   \label{fitting}
\end{equation}
where 
\begin{equation}
    \eta = \nH Z^{0.4}\chi^{-1.1},
    \label{eta}
\end{equation}
$\chi \equiv \sqrt{\chico \chioh}$, 
$\Ao$ is the relative oxygen abundance with respect to hydrogen 
in the gas phase, and 
${\cal A}_\mathrm{O,ism}=3.2\times 10^{-4}Z$ corresponds to 
$\Ao$ shown in Table \ref{tab:abundance}.

We should note that Equation (\ref{fitting}) is based on the chemical reactions in regions 
where shielding effects are not important. 
However, 
by considering the shielding effect in $\chi$,
we will show Equation (\ref{fitting}) 
is applicable also in regions where shielding effects
are effective in Section \ref{sec:shielding}.

\subsubsection{Application of the Analytical Formula 
to Regions Where Shielding Effects Work}\label{sec:shielding}

The analytic formula $\nCOana$ at a given position depends on $\nH$, $Z$, and $\chi=\sqrt{\chico\chioh}$.
Although $\nH$ and $Z$ are given locally, $\chi$ is 
determined by the column densities integrated from sources of light.
The FUV radiation that destroys CO can be attenuated by dust grains, C$^0$, CO, and H$_2$.
In debris disks, the dust extinction is negligible.

Which of C$^0$, CO, and H$_2$ is more important 
for the shielding effects?
In each shielding effect, 
one can define the critical column density at which $\chico$ is halved. 
Their critical densities are 
\begin{eqnarray}
&&N(\mathrm{H_2})_\mathrm{shld} \sim  4\times 10^{20}~\psc,\nonumber \\
&&N(\mathrm{C^0})_\mathrm{shld} \sim 4\times 10^{16}~\psc \\
&& N(\mathrm{CO})_\mathrm{shld} \sim 10^{14}~\psc\nonumber.
\label{COsh}
\end{eqnarray}
$N(\mathrm{H_2})_\mathrm{shld}$ is given by the results in \citet{vanDishoeck1988};
we derive the H$_2$ column density at which the shielding factor 
becomes 0.5 by the linear interpolation using Table 5 of their paper at $\log_{10} N(\mathrm{CO})=0$ \citep[also see][]{Visser2009}. 
$N(\mathrm{C}^0)_\mathrm{shld}$ is given by $(\ln 2)\alpha_\mathrm{C}^{-1}$,
where $\alpha_\mathrm{C}=1.777\times 10^{-17}~\mathrm{cm}^2$ is 
the cross section of photo-ionization of C$^0$ \citep{Heays2017},  
because the shielding factor owing to C$^0$ is proportional to $\exp(-\alpha_\mathrm{C}N(\mathrm{C}^0))$. 
$N(\mathrm{CO})_\mathrm{shld}$ is taken from 
the CO column density at which the shielding factor is 0.5 at $\log_{10} N(\mathrm{H}_2)=0$
in Table 5 of \citet[][also see Equation (\ref{fshield})]{Visser2009}.

Let us estimate the importance of the shielding 
effect owing to H$_2$.
Considering the abundance ratio between hydrogen and carbon, 
one finds that the H$_2$ column density is expressed as 
$N(\mathrm{H}_2) = 3.8\times 10^3 Z^{-1} N(\mathrm{C}^0)$, 
where it is assumed that all hydrogen is in H$_2$ ($n(\mathrm{H}_2)=\nH/2$) 
and all carbon is in C$^0$ ($n(\mathrm{C}^0)=\nC$). 
By contrast, in order for the shielding effect owing to H$_2$ to be more important than 
the C$^0$ attenuation, $N(\mathrm{H}_2)$ should be larger than 
$10^4 N(\mathrm{C}^0)$, 
where the numerical factor $10^4$ corresponds to 
$N(\mathrm{H}_2)_\mathrm{shld}/N(\mathrm{C}^0)_\mathrm{shld}$.
This condition is not satisfied even for $Z=1$, 
and the shielding effect owing to H$_2$ is not important 
compared with the C$^0$ attenuation.
For simplicity, 
the shielding by H$_2$ is neglected in 
the analytic formula.
We should note that the photo-dissociation rate 
may be overestimated in significantly shielded regions with 
$N(\mathrm{H_2})>N(\mathrm{H_2})_\mathrm{shld}$.

In debris disks, C$^0$ and CO contribute mainly to a decrease in $\chico$.
The problem is that the C$^0$ and CO fractions both depend on local $\chico$,
which depends on their column densities.
In this section, we construct a 
procedure to determine the  
the spatial distributions of $\nCI$, $\nCO$, and $\chico$ consistently.


Before presenting a way to evaluate the 
shielding effects, 
we should mention about the analytic formulae for $\nCO$ and $\nCI$.
The analytic formula for $\nCO$ shown in 
Equation (\ref{fitting}) is not applicable directly 
to a situation where carbon nuclei are mostly in CO 
because $\nCOana$ increases without limit as 
$\eta$ increases. To address this issue, 
Equation (\ref{fitting}) is simply modified 
so that $\nCO_\mathrm{ana}$ approaches $\nC$ smoothly 
in $\eta\rightarrow \infty$ as follows:
\begin{eqnarray}
    && \frac{\nCOana}{\nC}  =  \nonumber \\
    &&  \left[1 + \left\{\frac{\Ao}{ {\cal A}_\mathrm{O,ism}} 
    \left(
        10^{-14} \eta^{1.8} + 6.0\times 10^{-11} \eta 
\right)\right\}^{-1}\right]^{-1}.
\label{nCOana}
\end{eqnarray}
This modification may be ad hoc, but it predicts the CO fractions consistent with those obtained from the PDR calculations
as will be shown in  Figure \ref{fig:COshield}.

An analytic formula for the C$^0$ fraction is presented. 
The C$^0$ fraction is determined by the balance between photo-ionization of 
C$^0$ and radiative recombination of C$^+$ \citep{Kamp2000} as follows:
\begin{equation}
    \frac{n(\mathrm{C}^0)_\mathrm{ana}}{\nC-\nCO_\mathrm{ana}} = \frac{2\xi+1-\sqrt{1+4\xi}}{2\xi},
\label{nCIana}
\end{equation}
where the denominator $\nC-\nCO_\mathrm{ana}$ guarantees that the sum of 
the C$^0$, C$^+$, and CO number densities is equal to $\nC$,
and $\xi$ is the ratio of the recombination to the ionization coefficients that is given by 
\begin{eqnarray}
    \xi &\equiv& \frac{k_\mathrm{rec}(n_\mathrm{C}-\nCO_\mathrm{ana})}
   {\alpha_\mathrm{C} \chico F_\mathrm{H}} \nonumber \\
   &=& 
   \chico^{-1}
   \left(\frac{\nC -\nCO_\mathrm{ana}}{11~\mathrm{cm}^{-3}}\right)
    \left( \frac{T}{100~\mathrm{K}} \right)^{-0.82}.
     \label{xi1}
\end{eqnarray}
We confirmed that Equation (\ref{nCIana}) reproduced the C$^0$ fractions obtained from 
the PDR calculations.

The normalized FUV flux $\chico$ is expressed 
in terms of the unattenuated 
value $\chi_\mathrm{co,thin}$ as follows:
 \begin{equation}
     \chico = \chi_\mathrm{co,thin} f_\mathrm{shield}(N(\mathrm{C}^0)_\mathrm{ana},\NCO_\mathrm{ana}),
     \label{chico_shield}
 \end{equation}
 where $f_\mathrm{shield}$ corresponds to the shielding factor, which 
 depends on the C$^0$ and CO column densities ($N(\mathrm{C^0})_\mathrm{ana}$ 
 and $N(\mathrm{CO})_\mathrm{ana}$) computed from 
 $n(\mathrm{C}^0)_\mathrm{ana}$ and $\nCOana$, respectively.
 As the shielding factor, 
 we adopt the following expression,
 \begin{equation}
     f_\mathrm{shld} = e^{ - \alpha_\mathrm{C} N(\mathrm{C}^0)_\mathrm{ana} }
     \left[  1 + 
     \left(\frac{N(\mathrm{CO})_\mathrm{ana}}{10^{14}~\mathrm{cm}^{-2}} \right)^{0.6} \right]^{-1},
     \label{fshield}
 \end{equation}
 where the first factor on the right-hand side corresponds to the C$^0$ attenuation, 
 and the second factor is a fitting function of the self-shielding factor at $N(\mathrm{H}_2)=0$ tabulated in \citet{Visser2009}.

The spatial distributions of $\chico$, $n(\mathrm{C}^0)_\mathrm{ana}$, and $\nCOana$ are 
determined consistently in a iterative manner by using 
Equations (\ref{nCOana}), (\ref{nCIana}), and (\ref{chico_shield}).

It is useful to derive $\nH$ required to produce a specific value of the CO fraction.
In the high density limit where Path$_\mathrm{OH}$ is dominated, since the second term in the right-hand side of Equation (\ref{fitting}) is negligible,
Equation (\ref{fitting}) is solved for $\nH$ as follows:
\begin{eqnarray}
    n_\mathrm{H,req} &=& 6\times 10^7~\pcc~Z^{-0.4}\chi^{1.1} \nonumber \\
    && \;\;\;\;
    \times \left( \frac{\Ao}{ {\cal A}_\mathrm{O,ism}} \right)^{-0.56} 
    \left( \frac{n(\mathrm{CO)}}{\nC} \right)^{0.56},
    \label{nHconstrained}
\end{eqnarray}
which is valid when
\begin{equation}
    n_\mathrm{H,req} > 5.3\times 10^4~\pcc~Z^{-0.4}\chi^{1.1}.
    \label{nHreq}
\end{equation}
Equation (\ref{nHconstrained}) is rewritten as 
\begin{eqnarray}
    n_\mathrm{C,req} &=& 8\times 10^3~\pcc Z^{0.6}\chi^{1.1} \nonumber \\
    && \;\;\;\;
    \times \left( \frac{\Ao}{ {\cal A}_\mathrm{O,ism}} \right)^{-0.56} 
    \left( \frac{n(\mathrm{CO})}{\nC} \right)^{0.56},
    \label{nCreq}
\end{eqnarray}
which corresponds to the carbon nucleius density 
required to reproduce a specific value of $\nCO/\nC$.
From Equation (\ref{nCreq}), an important conclusion can be drawn that $\nCO/\nC$ decreases with increasing $Z$
for fixed $\nC$ and $\chi$.
In other words, the upper limit of the CO fraction is obtained at $Z=1$.

\subsubsection{Comparison of the CO Fractions with Those Predicted from the Analytical Formula}\label{sec:compana}



Figure \ref{fig:COshield} compares 
the results of the PDR calculations 
with the predictions from the analytic formula. 
$\nCOana$ reproduces the spatial profiles of $\nCO$ reasonably well 
from the unattenuated regions ($\NC\ll 10^{17}~\psc$) to 
the shielded regions ($\NC\gtrsim 10^{17}~\psc$)
for all the parameter sets.

For lower $\nC$ shown in Figures \ref{fig:COshield}a 
and \ref{fig:COshield}c, the CO fractions begin to 
increase around $\NC\sim 10^{16-17}~\psc$, regardless of $Z$ and $\chi$.
By contrast, Figures \ref{fig:COshield}b and \ref{fig:COshield}d show that 
the CO fractions with $Z=1$ begin to increase at lower $\NC$.
This comes from the fact that CO self-shielding only works if the CO fraction is high enough; otherwise, C$^0$ attenuation 
becomes more important.

A critical CO fraction above which 
CO self-shielding effect works is given by 
\begin{equation}
    \left(\frac{\nCO}{\nC}\right)_\mathrm{cri} \sim \frac{\NCO_\mathrm{shld}}{N(\mathrm{C}^0)_\mathrm{shld}}
    \sim  3\times 10^{-3}.
    \label{COshield_cri}
\end{equation}
The horizontal dashed lines in Figure \ref{fig:COshield} 
correspond to $(\nCO/\nC)_\mathrm{cri}$. 
In order to illustrate the effect of the 
CO self-shielding 
in Figure \ref{fig:COshield}, 
we overplot the profiles of $\nCO_\mathrm{ana}$ without 
considering the CO self-shielding effect in Equation (\ref{fshield}) by 
the thick dashed lines.
It is clearly seen that the CO self-shielding effect 
increases the CO fractions at
$N_\mathrm{C}<N(\mathrm{C}^0)_\mathrm{shld}$
only when $\nCO$ exceeds 
$\nCO_\mathrm{cri}$.

The importance of C$^0$ attenuation in CO formation has been pointed 
out by \citet{Kral2019}. 

\subsection{Parameter Survey}\label{sec:parametersurvey}

In Section \ref{sec:analyticmodel}, we developed 
the analytic formula which reproduces the results of the PDR calculations reasonably 
well when the H$_2$ fraction is too small to affect the CO formation.
By contrast, in Section \ref{sec:plane}, 
we found that the models with 
lower $T_\mathrm{chem}$ and/or higher $\fracdg$ 
yield a sufficient amount of H$_2$ to  accelerate the CO formation. 
In this section, we investigate how the CO fractions depend on $\fracdg$, $T_\mathrm{chem}$, and 
$\NC$ for various $n_\mathrm{C}$, $Z$, and $\chico$.

\begin{figure*}[htpb]
        \centering
        \includegraphics[width=18.0cm]{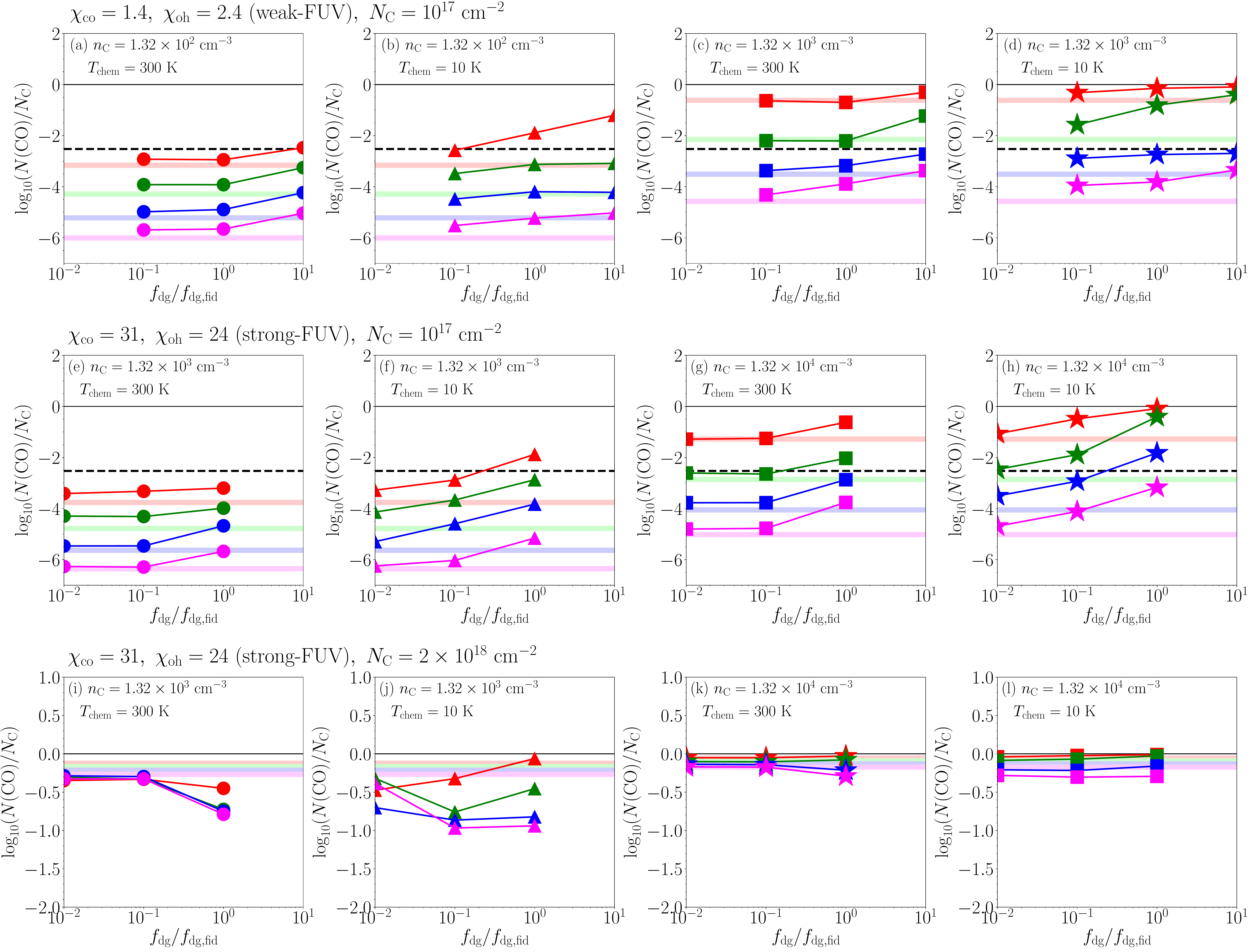}
        \caption{
        CO column density ratios $N(\mathrm{CO})/N_\mathrm{C}$ as a function of $\fracdg/\fracdgfid$ 
        for ({\it top panels}) the weak-FUV models with $\NC=10^{17}~\psc$,
        ({\it middle panels}) the strong-FUV models with $\NC=10^{17}~\psc$, and 
        ({\it bottom panels}) the strong-FUV models with $\NC=2\times 10^{18}~\psc$. 
        The results with $2\times 2$ combinations of 
        $\nC=(1.32\times 10^2~\pcc, 1.32\times 10^3~\pcc)$
        and $T_\mathrm{chem}=(300~\mathrm{K},10~\mathrm{K})$ for weak-FUV and 
        $\nC=(1.32\times 10^3~\pcc, 1.32\times 10^4~\pcc)$
        and $T_\mathrm{chem}=(300~\mathrm{K},10~\mathrm{K})$ for strong-FUV. 
        In each panel, the color represents the gas metallicities,
        (red) $Z=1$, (green) $Z=10$, (blue) $Z=10^2$, and (magenta) $Z=10^3$.
        The horizontal thick lines correspond to the predictions from $\nCO_\mathrm{ana}$ with 
        four different $Z$ ($Z=1,10,10^2,10^3$).
        The horizontal dashed black lines show the critical CO fraction $(\nCO/\nC)_\mathrm{cri}$ (Equation (\ref{COshield_cri}))
        above which the CO is shielded mainly by self-shielding.
        }
\label{fig:COfit}
\end{figure*}

\subsubsection{The CO Fractions at $\NC = 10^{17}~\psc$}\label{sec:CO1e17}

First, we investigate the CO fractions at $\NC=10^{17}~\psc$, which 
is the mid-plane column density inferred from the observational results 
of $\beta$ Pictoris
(Section \ref{sec:modelparameter}).
At this column density, CO is not shielded by C$^0$ attenuation significantly while
CO self-shielding can work if the CO fraction is sufficiently high (Section \ref{sec:compana}).
For reference, we also investigate the CO fractions at 
$\NC=10^{17}~\psc$ for strong-FUV.
The top and middle panels of Figure \ref{fig:COfit} 
show the CO fractions as a function of $\fracdg/\fracdgfid$
for various $\nC$, $T_\mathrm{chem}$, and $\chico$.
Overall, as $T_\mathrm{chem}$ decreases  and/or 
$\fracdg$ increases,
the CO fractions increase and deviate from the predictions from the analytic formula.


First, we investigate the results with $T_\mathrm{chem}=300$~K 
(Figures \ref{fig:COfit}a, \ref{fig:COfit}c, \ref{fig:COfit}e, and 
\ref{fig:COfit}g).
There is a critical $\fracdg/\fracdgfid$ ($(\fracdg/\fracdgfid)_\mathrm{cri}$) 
above which the CO formation is accelerated.
$(\fracdg/\fracdgfid)_\mathrm{cri}$ 
depends on the parameters. 
For the fiducial weak-FUV models ($\nC=1.32\times 10^2~\pcc$),
$(\fracdg/\fracdgfid)_\mathrm{cri}$ is around 1, 
regardless of $Z$ (Figure \ref{fig:COfit}a).
For the high-density weak-FUV models ($\nC=1.32\times 10^3~\pcc$), $(\fracdg/\fracdgfid)_\mathrm{cri}$ depends on $Z$ (Figure \ref{fig:COfit}c).
The $Z\ge 10^2$ models show acceleration of the CO formation at $\fracdg/\fracdgfid=1$ while 
$(\fracdg/\fracdgfid)_\mathrm{cri}$ is around 1 for $Z\le 10$.

The $\fracdg$ dependence of the CO fractions for the fiducial 
strong-FUV models ($\nC=1.32\times 10^3~\pcc$) is 
similar to that for the high-density weak-FUV models ($\nC=1.32\times 10^3~\pcc$) (see Figures \ref{fig:COfit}c and \ref{fig:COfit}e);
enhancement of the CO fractions is seen only for $Z\gtrsim 10^2$ when $\fracdg/\fracdgfid$ is increased from 0.1 to 1.
For the high-density strong-FUV model ($\nC=1.32\times 10^4~\pcc$), $(\fracdg/\fracdgfid)_\mathrm{cri}\sim 0.1$ for all $Z$.

The short summary is that $(\fracdg/\fracdgfid)_\mathrm{cri}\sim 1$ for $\nC=1.32\times 10^2~\pcc$,
$(\fracdg/\fracdgfid)_\mathrm{cri}\sim 0.1$ for $\nC=1.32\times 10^4~\pcc$, and 
the models with $\nC=1.32\times 10^3~\pcc$ show the intermediate behavior where 
$(\fracdg/\fracdgfid)_\mathrm{cri}$ takes values between 0.1 and 1 depending on $Z$.
The dependence of $\chico$ on $(\fracdg/\fracdgfid)_\mathrm{cri}$ appears to be weak.

\begin{figure}[htpb]
        \centering
        \includegraphics[width=8.0cm]{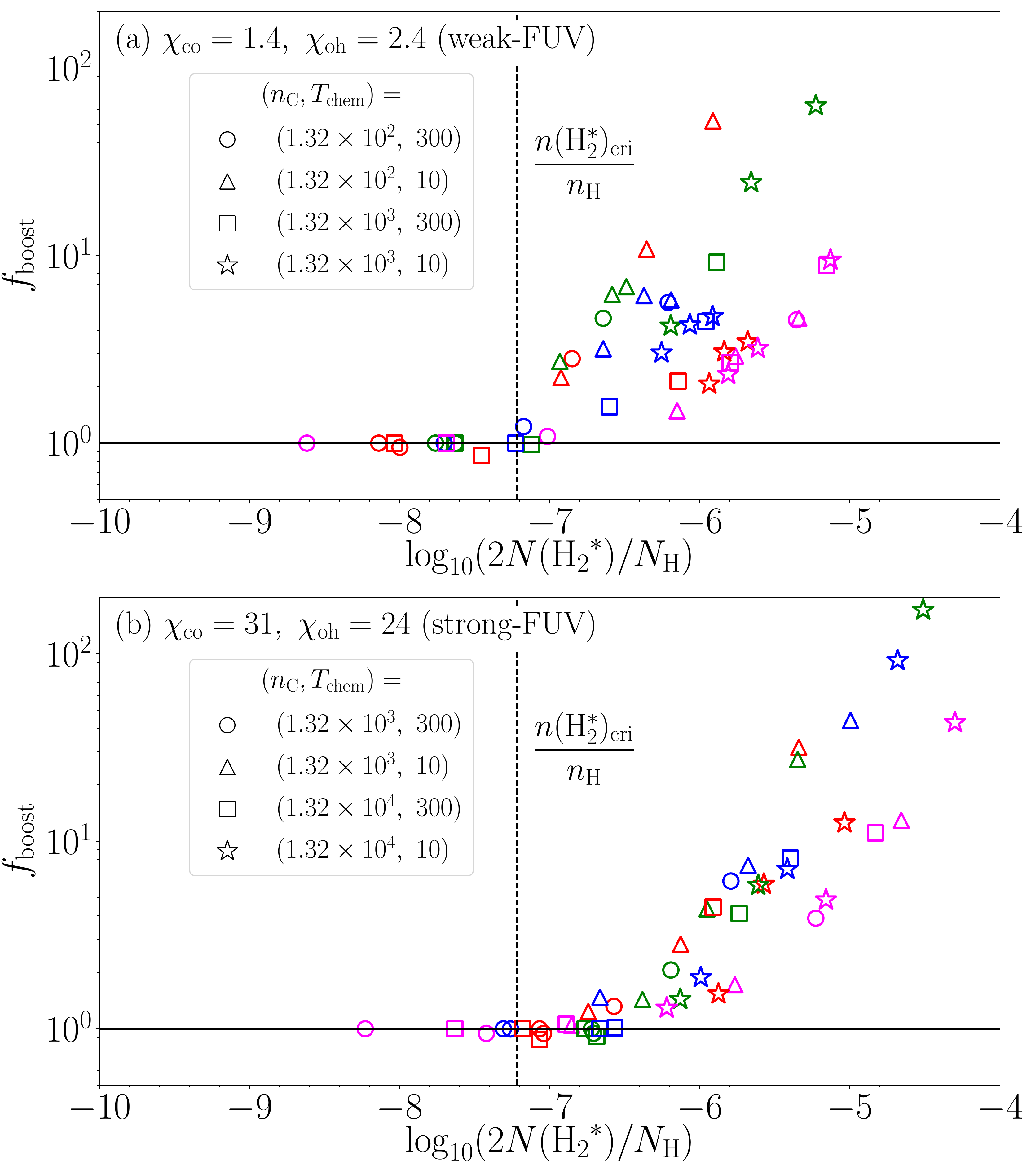}
        \caption{
        Scatter plots of $f_\mathrm{boost}$ versus $2N(\mathrm{H}^*_2)/\NH$, where 
        $f_\mathrm{boost}$ shows the degree of the enhancement of the CO fraction owing to H$^*_2$
        for (a) weak-FUV and (b) strong-FUV.
        The column densities are measured at $\NC=10^{17}~\psc$.
        The ranges of the parameters ($\nC, T_\mathrm{chem}, Z, \fracdg$) are shown in Table \ref{tab:param}.
        The data points are distinguished by $\nC$, $T_\mathrm{chem}$, and $Z$.
        The shapes of the markers represent the difference in ($\nC$, $T_\mathrm{chem}$).
        The colors of the markers represent different gas metallicities,
        (red) $Z=1$, (green) 10, (blue) 10$^2$, and (magenta) $10^3$.
        The data points with different $\fracdg$ are illustrated without distinction, but 
        $2N(\mathrm{H^*_2})/\NH$ increase with $\fracdg$.
        The vertical dashed black lines correspond to $2n(\mathrm{H_2^*})_\mathrm{cri}/\nH$ defined in Equation (\ref{nH2cri}).
	 }
\label{fig:H2CO}
\end{figure}

A decrease in $T_\mathrm{chem}$ reduces 
$(\fracdg/\fracdgfid)_\mathrm{cri}$ significantly.
Figures \ref{fig:COfit}b, \ref{fig:COfit}d, \ref{fig:COfit}f, 
and \ref{fig:COfit}h show that 
$(\fracdg/\fracdgfid)_\mathrm{cri}<0.1$ for weak-FUV and 
$(\fracdg/\fracdgfid)_\mathrm{cri}\sim 10^{-2}$ for strong-FUV.
The CO fractions monotonically increase with 
$\fracdg/\fracdgfid$ for most cases.

We should note that increases in the CO fractions appear to saturate for 
the weak-FUV models with ($n_\mathrm{C}=1.32\times 10^2~\pcc$, $\fracdg\ge \fracdgfid$, $Z > 1$, $T_\mathrm{chem}=10~$K) and 
($n_\mathrm{C}=1.32\times 10^3~\pcc$, $\fracdg\ge \fracdgfid$, $Z > 10$, $T_\mathrm{chem}=10~$K). 
This is because the H$_2$ fractions are close to unity in these models, and 
the enhancement of the H$_2$ formation rate does not lead to further acceleration of the CO fractions.

The CO fractions at $\NC=10^{17}~\psc$ are affected by CO self-shielding when the CO fractions exceed 
$(\nCO/\nC)_\mathrm{cri}$ (Equation (\ref{COshield_cri})), which is shown by the horizontal dashed black lines.
When $\nCO/\nC>(\nCO/\nC)_\mathrm{cri}$, a small increase in $\nCO/\nC$ for $\NC<10^{17}~\psc$ results in 
a large increase in $\NCO/\NC$ at $\NC=10^{17}~\psc$.
This effect is clearly seen in 
Figures \ref{fig:COfit}b and \ref{fig:COfit}d. 

What chemical reactions contribute to the acceleration of the CO formation?
We found that vibrationally excited H$_2$ (hereafter H$^*_2$) plays an important role. 
As a result, different reaction paths to CO are activated than the chemical reactions shown in 
Figure \ref{fig:COnetwork}.
It starts from $\mathrm{C^+ + H_2\rightarrow CH^+ + H}$ 
which is known to be endothermic by $\sim 4300~$K.
The internal energy of H$^*_2$ is available to 
overcome its energy barrier. 
The reaction between C$^+$ and H$^*_2$ with $v>0$ 
proceeds at almost the Langevin collision rate 
($k_\mathrm{cp,h2*}=1.6\times 10^{-9}~\mathrm{cm}^3~\mathrm{s}^{-1}$) \citep{Hierl1997}.

The efficiency of 
the mechanism is investigated in \citet{Zanchet2013} and \citet{Herrez-Aguilar2014}.
In PDRs of the ISM, the importance of H$^*_2$ was also 
pointed out in many references \citep{Agndez2010,Goicoechea2016,Joblin2018,Veselinova2021,Goicoechea2022}.


How much H$^*_2$ is required to accelerate CO formation?
Figure \ref{fig:H2CO} shows the scatter plots of $2N(\mathrm{H}^*_2)/\NH$ versus a boost factor $f_\mathrm{boost}$, which 
is defined as the CO fraction divided by that at the H$_2$-poor environments where 
only $\fracdg$ is changed to $0.1\fracdgfid$ for weak-FUV 
and $0.01\fracdgfid$ for strong-FUV, keeping the other parameters unchanged.

Figure \ref{fig:H2CO} show that 
$f_\mathrm{boost}$ is correlated with 
$2N(\mathrm{H}^*_2)/\NH$ for both weak-FUV and strong-FUV, 
indicating that H$^*_2$ accelerates the CO formation.
$f_\mathrm{boost}$ is larger than unity only when $2N(\mathrm{H}_2^*)/\NH$ exceeds a critical 
H$_2^*$ fraction of $\sim 10^{-7}$ although there are large scatters.

The critical H$^*_2$ fraction can be roughly understood as follows.
A necessary condition affecting the CO formation by H$^*_2$ is that 
the reaction rate of $\mathrm{C^+ + H_2^*\rightarrow CH^+ + H}$ 
must be larger than that of $\mathrm{C^++H \rightarrow CH^+ + h\nu}$. 
The condition becomes 
\begin{eqnarray}
n(\mathrm{H}_2^*) > 
    n(\mathrm{H}_2^*)_\mathrm{cri} &=& \frac{k_\mathrm{cp,h}}{k_\mathrm{cp,h2*}}\nH \nonumber \\
   & \sim & 3\times 10^{-8}\left(\frac{T}{50~\mathrm{K}}\right)^{-0.42}\nH,
    \label{nH2cri}
\end{eqnarray}
where  $k_\mathrm{cp,h} =
2.29\times 10^{-17}(T/300)^{-0.42}~\mathrm{cm^3~s^{-1}}$ is the reaction rate of 
$\mathrm{C^+ + H\rightarrow CH^+ + h\nu}$.
Figures \ref{fig:H2CO} show that 
$2N(\mathrm{H_2^*})/\NH > 2n(\mathrm{H_2^*})_\mathrm{cri}/\nH$
can distinguish whether to accelerate the CO formation or not. 

The correlation between $f_\mathrm{boost}$ 
and $2N(\mathrm{H_2^*})/\NH$ 
depends on $Z$; the models with lower $Z$ 
tend to show larger $f_\mathrm{boost}$
at a fixed $2N(\mathrm{H_2^*})/\NH$.
This is because 
for smaller $Z$, the efficient 
formation of CH$^+$ activates more reaction pathways involved by hydrogen to produce CO 
since $\nH$ increases with decreasing $Z$ at a given $\nC$.

\subsubsection{The CO Fractions at $\NC = 2\times 10^{18}~\psc$}\label{sec:CO2e18}

We investigate the CO fractions for strong-FUV at $\NC=2\times 10^{18}~\psc$, which 
is the mid-plane column density inferred from the observational results 
of 49 Ceti (Section \ref{sec:modelparameter}).
At this column density, CO is shielded significantly.

Interestingly the CO fractions 
at $\NC=2\times 10^{18}~\psc$ show the opposite
$\fracdg$-dependence from that at $\NC=10^{17}~\psc$.
For $T_\mathrm{chem}=300~$K, 
increases in $\fracdg$ reduce the CO fractions 
(Figures \ref{fig:COfit}i and \ref{fig:COfit}k).
The production of CH$^+$ through the reaction between C$^+$ and H$^*_2$ 
does not contributes to the CO formation because the  C$^0$ attenuation 
makes the C$^+$ fractions extremely low. 
That is why efficient formation of H$_2$ no longer promotes the CO formation.
Conversely, the presence of H$_2$ negatively affects CO formation.
In the chemical network shown in Figure \ref{fig:COnetwork}, 
an increase in the amount of H$_2$ 
reduces the amount of available H atoms, 
leading to a decrease in the CO formation rate.

A similar behavior is seen for $T_\mathrm{chem}=10~$K.
As long as $\fracdg\le 0.1\fracdgfid$, 
increasing $\fracdg$ reduces the amount of CO.
When $\fracdg$ is increased from $0.1\fracdgfid$ to $\fracdgfid$, 
the CO fractions turn to increase 
in Figures \ref{fig:COfit}j and \ref{fig:COfit}l.
This is because most hydrogen exists as H$_2$ for $\fracdg=\fracdgfid$. 
The chemical reactions shown in Figure \ref{fig:COnetwork} 
no longer work, and different chemical 
reactions associated with H$_2$ become 
important to form CO.


\section{Discussion}\label{sec:discuss}

\subsection{ Predictions of the Spatial Distributions of CO in a Disk Structure}\label{sec:COdiskana}

The Meudon PDR code is not designed to investigate the chemical and thermal structure of disks.
However, in Section \ref{sec:analyticmodel}, we developed the numerical procedure to determine the CO fractions 
on the basis of the findings 
of the plane-parallel PDR calculations.
In this section,
we present the method to derive the spatial distributions of 
$\nCIana$ and $\nCOana$ in a given disk structure using a similar method as shown in Section \ref{sec:shielding}.  


\begin{figure}[htpb]
        \centering
        \includegraphics[width=8cm]{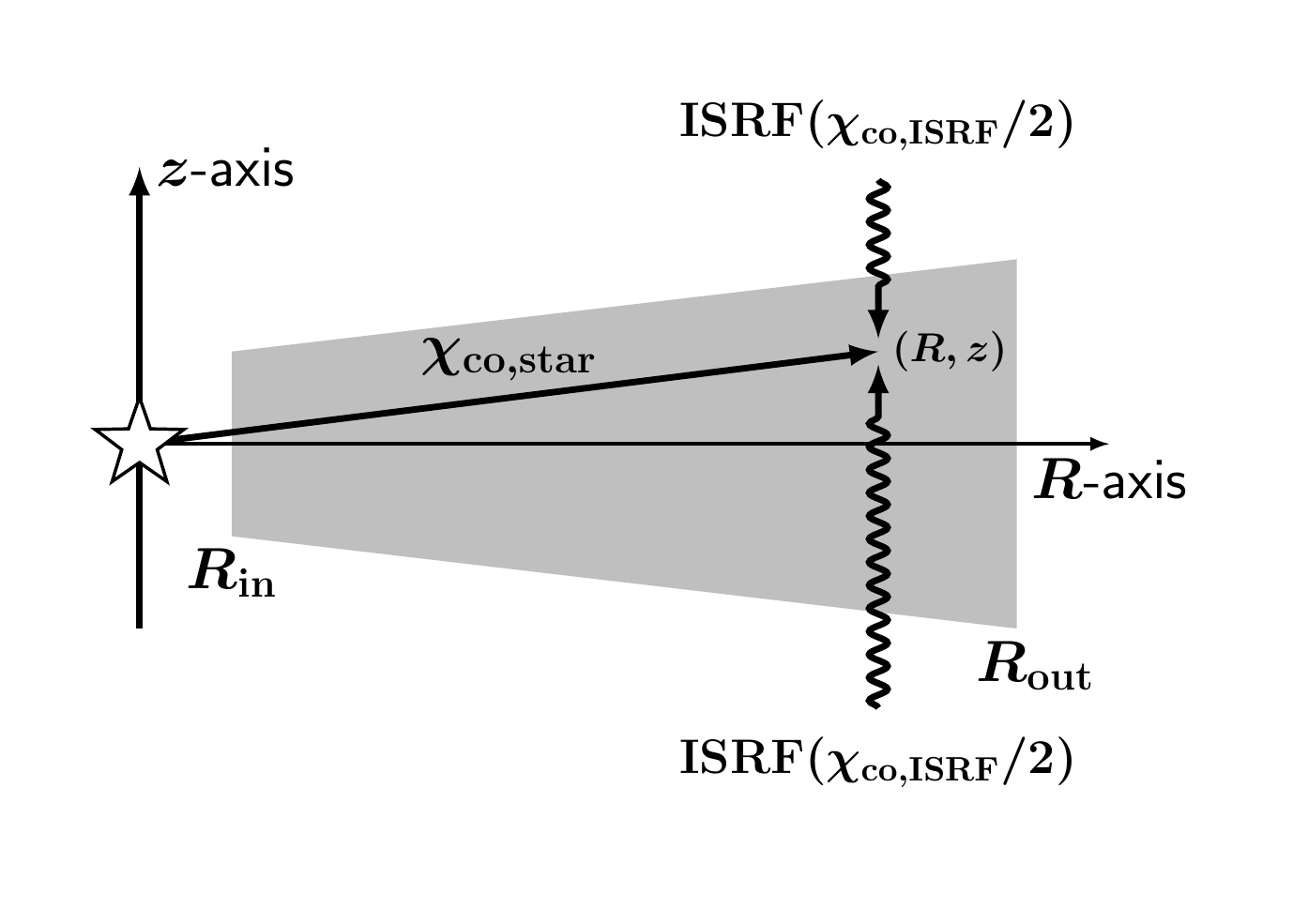}
        \caption{
            Schematic picture of our setting 
            where the three rays are considered.
	}
\label{fig:obs}
\end{figure}

Figure \ref{fig:obs} shows the disk structure considered in this paper.
The disk extends from $R=R_\mathrm{in}$ to $R_\mathrm{out}$, 
where the cylindrical coordinate $(R,z)$ is used and the central star locates at the origin.
The density distribution of carbon nuclei is denoted as $n_\mathrm{C}(R,z)$,
and $Z$ and $\fracdg$ are assumed to be uniform throughout the disk.
At a certain position $(R,z)$, 
the local radiation field is determined by the three rays, 
the stellar radiation that 
penetrates the disk from the inner edge to $(R,z)$ 
and the ISRF that propagates vertically 
from above and below the disk toward $(R,z)$.


Applying Equation (\ref{chico_shield}) into the cases with the disk geometry, 
one obtains  $\chico(R,z)$ as follows:
\begin{eqnarray}
    \chico(R,z) &=& \chi_\mathrm{co,star} \left( \frac{\sqrt{R^2+z^2}}{r_0} \right)^{-2} 
    f_{\mathrm{shield},r}(R,z) \nonumber  \\
    &+& \frac{\chi_\mathrm{co,ISRF}}{2} f_{\mathrm{shield},+z}(R,z) \nonumber  \\
    &+& \frac{\chi_\mathrm{co,ISRF}}{2} f_{\mathrm{shield},-z}(R,z), 
    \label{chicoRz}
\end{eqnarray}
where $\chi_\mathrm{co,star}$ and $\chi_\mathrm{co,ISRF}$ are 
the normalized unattenuated 
FUV fluxes of the stellar radiation at a reference radius of $r_0=50~\mathrm{au}$ and 
the ISRF, respectively. 
The geometrical dilution is considered in the first term on the right-hand side of Equation (\ref{chicoRz}).
A shielding factor is defined for each of the three rays;
$f_{\mathrm{shield},r}$ is the shielding factor of the stellar radiation, and 
$f_{\mathrm{shield},+z}$ ($f_{\mathrm{shield},-z}$) 
is the shielding factor of the ISRF coming from below (above) the disk.
They are computed using Equation (\ref{fshield}), but the column densities 
$\NCO_\mathrm{ana}$ and $\NCI_\mathrm{ana}$ 
are computed by integrating 
$\nCO_\mathrm{ana}$ and $\nCI_\mathrm{ana}$ 
along the corresponding rays, respectively.

The normalized FUV flux $\chioh$ is given by 
\begin{eqnarray}
    \chi_\mathrm{oh}(R,z) &=& \chi_\mathrm{oh,star} \left( \frac{\sqrt{R^2+z^2}}{r_0} \right)^{-2},
    \label{chiohRz}
\end{eqnarray}
where $\chi_\mathrm{oh,star}$ 
is the normalized unattenuated 
FUV flux of the stellar radiation at $r_0=50~\mathrm{au}$. 
Shielding effects are not effective in $\chioh$ 
as mentioned in Section \ref{sec:fuv}


From Equations  (\ref{nCIana}), (\ref{nCOana}), (\ref{chicoRz}), and (\ref{chiohRz}),
$\nCIana(R,z)$, $\nCOana(R,z)$, and $\chico(R,z)$
are determined consistently in an iterative manner.

\subsection{Comparison with Observations}\label{sec:observation}

Our analytical model shown in Section \ref{sec:COdiskana} is 
applied to the debris disks around $\beta$ Pictoris and 49 Ceti
in Sections \ref{sec:betaPic} and \ref{sec:49Ceti}, respectively.

Although $\nCIana(R,z)$ and $\nCOana(R,z)$ are obtained throughout the disk,
additional PDR calculations are conducted at various radii along the mid-plane considering
the stellar radiation and vertical ISRF.
There are two reasons for this.
One is to verify whether $\nCIana$ and $\nCOana$ reproduce 
the results of the PDR calculations with $T_\mathrm{chem}=300~$K.
The other is to investigate how the H$_2$ formation accelerated 
by setting $T_\mathrm{chem}=10$~K affects the CO formation.
Section \ref{sec:pdrdisk} will describe how the PDR calculations 
are conducted taking into account the disk geometry.

\subsubsection{PDR Calculations Using Disk Geometry and Structure}
\label{sec:pdrdisk}

In the PDR calculation at $(R=R_i,z=0)$, supposing that 
$I_\mathrm{star}(\lambda)$ is the stellar radiation intensity 
and $I_\mathrm{ISRF}(\lambda)$ is the ISRF intensity, 
the mean intensity $J(\lambda)$ at $(R_i,0)$ is given by 
\begin{eqnarray}
    J(\lambda) &=& 
        I_\mathrm{star}(\lambda)W( R_i)
        f_{\mathrm{shield},r}(R_i,0) \nonumber \\
        &+& I_\mathrm{ISRF}(\lambda)
        f_{\mathrm{shield},+z}(R_i,0)
    \label{radPDR}
\end{eqnarray}
for $\lambda \le 1100~$\AA, 
where $W(R_i)$ is the geometrical dilution 
factor at $R=R_i$, and 
we use the fact that $f_{\mathrm{shield},-z}(R_i,0) = f_{\mathrm{shield},+z}(R_i,0)$.
For the spectrum with $\lambda > 1100~$\AA, 
the shielding factors in Equation (\ref{radPDR}) are set to unity.


For each $R_i$, we perform the PDR calculation with 
$\nC(R_i,0)$, $Z$, $\fracdg/\fracdgfid$, and 
the radiation field shown in Equation (\ref{radPDR}). 
Since the geometrical dilution and shielding effects are already taken into 
account in the incident radiation, 
the chemical abundances at the optically thin limit ($N_\mathrm{C}=10^{14}~\psc$) 
are considered as those at the position $(R_i,0)$.

\subsubsection{$\beta$ Pictoris}\label{sec:betaPic}


    On the basis of the disk models shown in \citet{Cataldi2018}, 
    we here consider a ring with a uniform carbon nucleus vertical column density.
    With the C$^0$ and C$^+$ line data 
    \citep{Cataldi2014,Cataldi2018},
    \citet{Cataldi2018} derive the best fit parameters 
    of $R_\mathrm{in}=50~\mathrm{au}$ 
    and $R_\mathrm{out}=120~$au. 
    The stellar parameters of the central star 
    adopted here are the same 
    as those of the A5V star shown in Section \ref{sec:fuv}.
    The scale height is 
    $h(R) = c_\mathrm{s}(R)/\Omega(R)$, 
    where $c_\mathrm{s}$ is the sound speed with 
    the constant temperature 
    $T=75~\mathrm{K}$ and $\Omega(R) = \sqrt{G1.75M_\odot/R^3}$.


    Since ${\cal N}_\mathrm{C}\sim 10^{17}~\psc$ for $\beta$ Pictoris
    (Section \ref{sec:modelparameter}), 
    an average mid-plane carbon nucleus number density 
    is estimated to be 
    $\sim {\cal N}_\mathrm{C}/
    (R_\mathrm{out}-R_\mathrm{in})\sim 100~\pcc$.
    In order to investigate how the CO fractions depend on the mid-plane density,  
    we consider two different values of
    $n_\mathrm{C,mid0}=\nC(R_\mathrm{in},0)$ at the inner edge,
    $n_\mathrm{C,mid0}=75~\pcc$ (low-density model) 
    and $n_\mathrm{C,mid0}=190~\pcc$ (high-density model) as listed in Table \ref{tab:betaPic}.
    
    \begin{table}
    \begin{center}
    \begin{tabular}{|l|c|c|}
        \hline
        & low-density & high-density \\
        \hline
        $n_\mathrm{C,mid0}$ [$\pcc$]$^{(1)}$ & 75 & 190  \\
        \hline
        ${\cal N}_\mathrm{C}$ [$\psc$]$^{(2)}$ & $4\times 10^{16}$ & $10^{17}$ \\
        \hline
        ${\cal N}_\mathrm{C\perp}$ [$\psc$]$^{(3)}$ & $1.8\times 10^{16}$ & $4.4\times 10^{16}$ \\
        \hline
        $f_\mathrm{dg}/f_\mathrm{dg,fid}$ & 5.0 & 2.0 \\
        \hline
    \end{tabular}
    \end{center}
    \caption{
    List of the model parameters for $\beta$ Pictoris.
    $^{(1)}$The mid-plane carbon nucleus number densities at $R=50~$au. 
    $^{(2)}$The mid-plane carbon nucleus column densities integrated from $R_\mathrm{in}$ to $R_\mathrm{out}$.
    $^{(3)}$The vertical carbon nucleus number densities with $Z=1$. 
    }
    \label{tab:betaPic}
    \end{table}

    The radial distribution of the mid-plane density is given by 
    \begin{equation}
        n_\mathrm{C}(R,0) = 75~\mathrm{cm}^{-3} 
        \left( \frac{ n_\mathrm{C,mid0}}{75~\mathrm{\pcc}} \right) 
        \left( \frac{R}{50~\mathrm{au}} \right)^{-3/2}.
        \label{nCbeta}
    \end{equation}
    where the mean molecular weight $\mu$ 
    is assumed to be spatially constant.

    The mid-plane column densities ${\cal N}_\mathrm{C}$ integrated from $R=R_\mathrm{in}$ to 
    $R_\mathrm{out}$ for both the models are 
    shown in Table \ref{tab:betaPic}.
    They correspond to the minimum and maximum values of the 
    observed C$^0$ column densities multiplied by two 
    \citep{Higuchi2017,Cataldi2018}, where 
    the factor two comes from the contribution from C$^+$.

     The vertical column density ${\cal N}_\mathrm{C\perp}$ is 
     determined by multiplying $n_\mathrm{C}(R,0)$ by
     the scale height, which depends on 
     $T$ and $\mu$. Although $\mu$ depends 
     on the chemical state of the species, for simplicity, 
     we assume that all gases are neutral atoms, and 
     $\mu$ is given by $
     1.27(1 + 6.6\times 10^{-3}Z)/
     (1 + 5.5\times 10^{-4}Z)$.
     Even when the gas is fully molecular, 
     our results do not change significantly because 
     $\mu$ will be about a factor of two larger and 
     the vertical column density decreases by about 30\%.
     
    

Referring to Equation (\ref{fdg_tau_Nc}), 
the dust-to-gas mass ratio is expressed in terms of 
${\cal N}_\mathrm{C}$ and $\tau_\mathrm{mid}$
as follows:
\begin{equation}
    \frac{\fracdg}{\fracdgfid} = 
   \left( \frac{{\cal N}_\mathrm{C}}{2\times 10^{17}~\psc} \right)^{-1}
    \left( \frac{\taumid}{10^{-2}} \right).
    \label{fdg_determ}
\end{equation}
From Equation (\ref{fdg_determ}), 
the dust-to-gas mass ratios are 
given by $\fracdg/\fracdgfid = 5.0$ for the 
low-density model and $2.0$ for the  high-density model, where  $\taumid=10^{-2}$ is used (Table \ref{tab:betaPic}).

\begin{figure}[htpb]
        \centering
        \includegraphics[width=9.0cm]{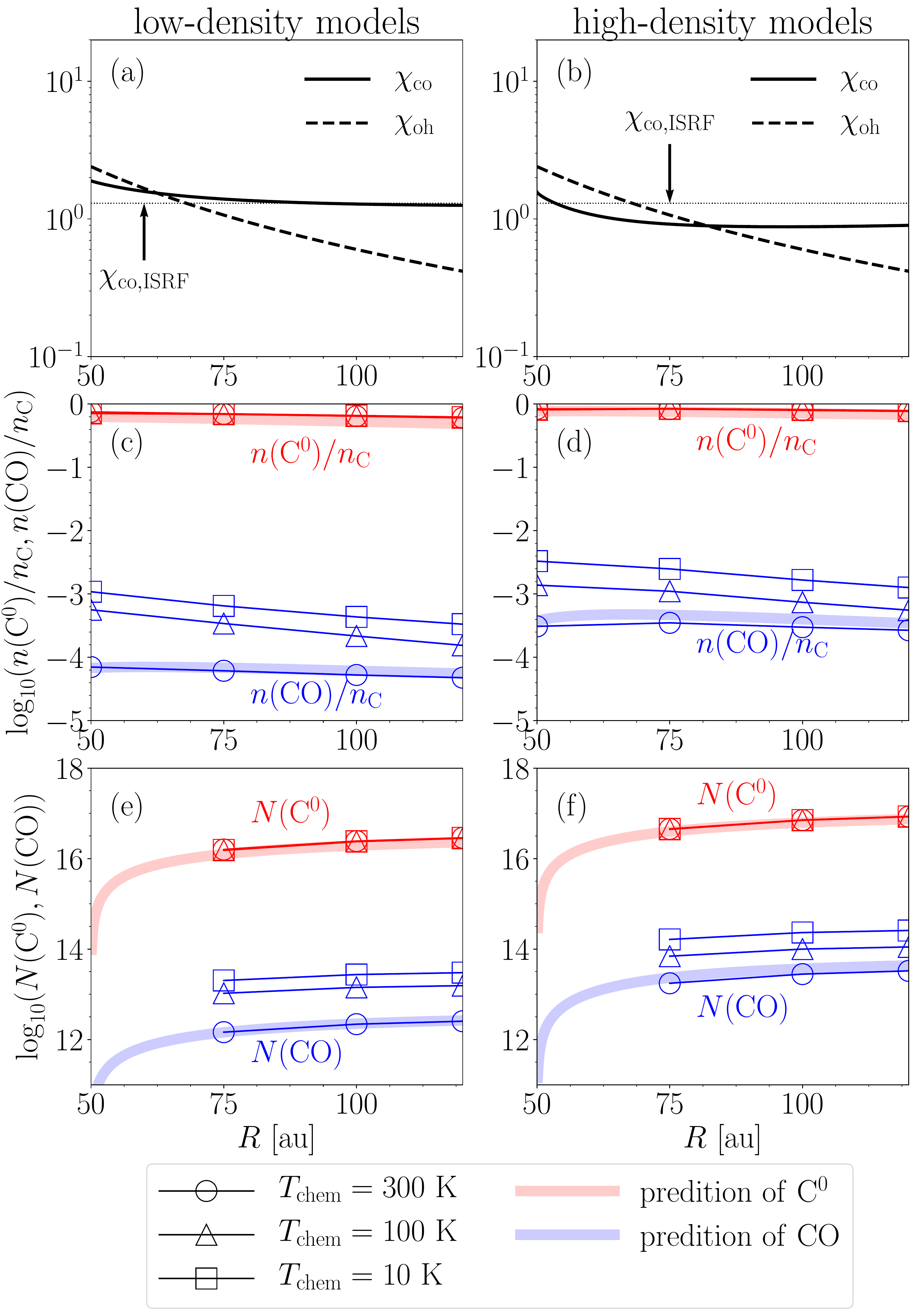}
        \caption{
             Results of the PDR calculations with $Z=1$ 
             for $\beta$ Pictoris.
             As the stellar radiation field, A5V star is adopted.
             The left and right columns show the results 
             of the low-density and high-density models, respectively.
             ({\it Top panels})  The normalized 
             UV fluxes (solid) $\chico$ and (dashed) $\chioh$. 
             The horizontal dotted lines correspond to $\chi_\mathrm{co,ISRF}$.
             ({\it Middle panels})
             The radial distributions of (red) $n(\mathrm{C}^0)/\nC$ and (blue) $\nCO/\nC$.
             The circles, triangles, and boxes correspond to the results 
             for $T_\mathrm{chem}=300~$K, $100~$K, and $10~$K, respectively.
             The C$^0$ and CO fractions predicted from 
             the analytic model (Section \ref{sec:COdiskana})
             are shown by the light red and blue solid lines, respectively.
             ({\it Bottom panels})
             The radial column densities of (red) atomic carbon and (blue) CO 
             integrated from the inner edge (50~pc) to $R$ 
             as a function of $R$.
}
\label{fig:betadisk}
\end{figure}

Using the analytical model 
presented in Section \ref{sec:COdiskana}, 
the spatial distribution of $\chico$ is obtained.
Figures \ref{fig:betadisk}a and \ref{fig:betadisk}b show 
$\chico$ at the mid-plane as a fuction of $R$. 
$\chico$ is almost constant because the ISRF gives 
dominant contribution in most radii except near the 
inner edge.
Since the vertical column density for the high-density model 
is around $N(\mathrm{C}^0)_\mathrm{shld}$, 
$\chi_\mathrm{co}$ is slightly attenuated.
By contrast, $\chioh$ at the mid-plane decreases with $R$ 
owing to the geometrical dilution.

    Before showing the results of the PDR calculations, 
    we present the approximate formula of the predictions from the analytic formula.
    Equation (\ref{nHreq}) can be rewritten as 
    $n_\mathrm{C,req}>7Z^{0.6}\chi^{-1.1}~\pcc$.
    Since $\nC\sim 100~\pcc$ and $\chi\sim O(1)$, 
    Equation (\ref{nCreq}) is valid for $Z\lesssim 10^2$.
    Equating the mid-plane density and the density required 
    to produce a given 
    CO fraction (Equation (\ref{nCreq})), 
    the CO fraction is expected to be 
    \begin{eqnarray}
        \left(\frac{\nCO}{\nC}\right)_\mathrm{\beta Pic} &=& 
        10^{-4}\left( \frac{n_\mathrm{C,mid0}}
        {75~\mathrm{cm}^{-3}} \right)^{1.8}
        \left( \frac{R}{50~\mathrm{au}} \right)^{-0.7} \nonumber \\
        & & \hspace{5mm} \times Z^{-1.1} 
        \chico^{-1} \left( \frac{ {\cal A}_\mathrm{O}}{ 
        {\cal A}_\mathrm{O,ism}} \right),
        \label{nCOpred_beta}
     \end{eqnarray}
     where Equation (\ref{chiohRz}) is used for $\chioh$,
     and $\chico$ is the attenuated local value.
     Integrating $\nCO$ over the disk extent with 
     Equations  (\ref{nCbeta}) and (\ref{nCOpred_beta}), 
     one obtains the predicted mid-plane CO column density,
    \begin{eqnarray}
        {\cal N}(\mathrm{CO})_\mathrm{\beta Pic} &=& 
        3\times 10^{12}~\psc\left( \frac{n_\mathrm{C,mid0}}
        {75~\mathrm{cm}^{-3}} \right)^{2.8} \nonumber \\
        & & \hspace{5mm} \times Z^{-1.1} 
        \langle \chico^{-1} \rangle \left( \frac{ {\cal A}_\mathrm{O}}{ 
        {\cal A}_\mathrm{O,ism}} \right),
        \label{NCOpred_beta}
     \end{eqnarray}
     where $\langle \chico^{-1}\rangle$ is the average of $\chico^{-1}$ weighted by 
     $R^{-2.2}$ over the disk extent.

The PDR calculations are conducted at four radii, $50~$au, $75~$au, 100~au, and 120~au.

%



    Firstly, the cases with $Z=1$ are considered.
    The results of the PDR calculations of the low-density
    and high-density models
    are shown in Figures \ref{fig:betadisk}c 
    and \ref{fig:betadisk}d, respectively.
     For $T_\mathrm{chem}=300$~K, the radial profiles 
    of $\nCO/\nC$ are consistent with the predictions from the analytic model (Section \ref{sec:COdiskana})
    for both the models 
    because H$_2$ formation is inefficient.
    Although $\fracdg/\fracdgfid$ is slightly larger than unity, 
    the effect of excited H$_2$ is limited 
    as shown in Figure \ref{fig:COfit}a.

    When $T_\mathrm{chem}$
    is decreased from $300$~K to $10~$K,
    the H$_2$ formation rate increases by about four 
    orders of magnitude at $T\sim 40~$K,
    making a large amount of hydrogen nuclei being molecular.
    As a result, H$^*_2$ increases the 
    CO fraction by about an order of magnitude 
    in both the models (Figures \ref{fig:betadisk}c and \ref{fig:betadisk}d).
     This is consistent with what we found in Figure \ref{fig:COfit}b.

    The mid-plane column densities of C$^0$ and CO integrated 
    from $R=R_\mathrm{in}$ to $R$ are shown in 
    Figures \ref{fig:betadisk}e and \ref{fig:betadisk}f as a function of $R$.
    After a rapid increase in $\NCO$ near the inner edge, 
    they become almost constant outside from $R\sim 75~$au.
    This clearly shows that the CO mid-plane column densities are determined near the inner edge, and 
    CO in the outer disk does not contribute to the total column density.

\begin{figure}[htpb]
        \centering
        \includegraphics[width=8cm]{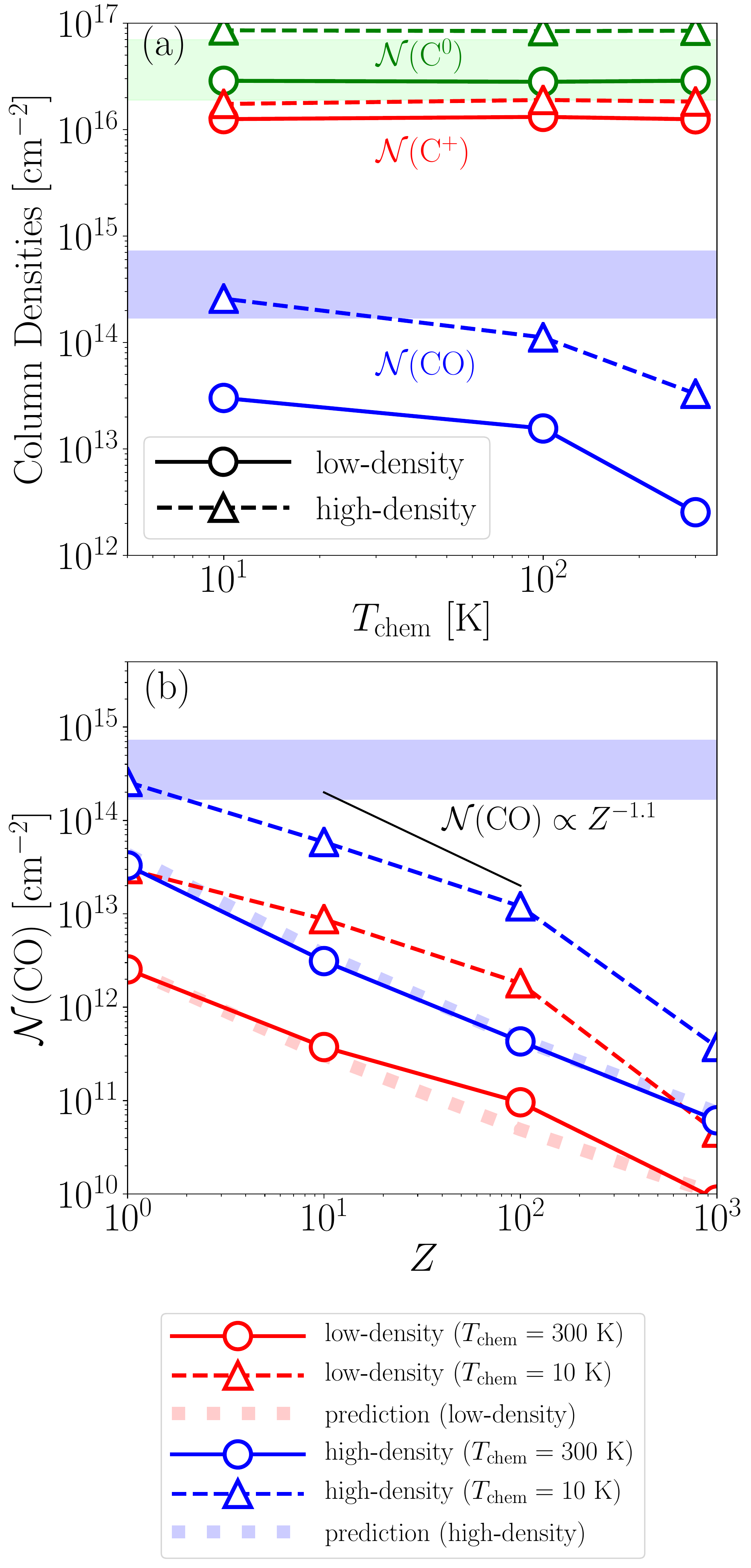}
        \caption{
            Results of the PDR calculations for $\beta$ Pictoris.
            (a) The mid-plane column densities of (red) C$^+$, (green) C$^0$, and (blue) CO 
            along the mid-plane as a function of $T_\mathrm{chem}$
            for (circles) the low- and (triangles) high-density models with $Z=1$.
            The green and blue rectangle regions indicate the observational constraints on 
            the C$^0$ and CO column densities obtained from 
            \citet{Higuchi2017,Cataldi2018} and \citet{Higuchi2017}, respectively.
            (b) The mid-plane CO column densities as a function of $Z$  for 
            (red) the low- and (blue) high-density models.
            The circles and triangles show the results with $T_\mathrm{chem}=300$~K 
            and $T_\mathrm{chem}=10$~K, respectively.
            The red and blue thick dashed lines correspond to the predictions from the analytic model (Section \ref{sec:COdiskana}) with 
            $T_\mathrm{chem}=300~$K and 10~K, respectively.
            The blue rectangle region is the same as that in Panel (a).
            As a reference, ${\cal N}(\mathrm{CO}) \propto Z^{-1.1}$ is plotted 
            (Equation (\ref{nCOpred_beta})).
}
\label{fig:beta_comp}
\end{figure}

In order to compare our results with the observational results, 
the mid-plane column  densities of C$^+$, C$^0$, and CO 
integrated from $R=R_\mathrm{in}$ to $R_\mathrm{out}$ are shown as 
a function of $T_\mathrm{chem}$ in Figure \ref{fig:beta_comp}a.
The C$^0$ column densities of the low- and 
high-density
models are both consistent with the observational results, 
and do not depend on $T_\mathrm{chem}$.
    
The CO column densities increase as 
the mid-plane gas density increases and/or 
$T_\mathrm{chem}$  decreases.
The high-density models produce 
about an order of magnitude larger 
CO column densities than the low-density model for all $T_\mathrm{chem}$.
This comes from the fact that 
${\cal N}(\mathrm{CO}) \propto n_\mathrm{C,mid0}^{2.8}$ 
(Equation (\ref{NCOpred_beta})).
The contribution of the attenuation of $\chico$ to 
an increase in ${\cal N}(\mathrm{CO})$ is limited since 
the vertical column density ${\cal N}_\mathrm{C,\perp}$ 
is comparable to or smaller than 
$N(\mathrm{C^0})_\mathrm{shld}$ (Table \ref{tab:betaPic}).
The CO column density of the high-density model with $T_\mathrm{chem}=300$~K
does not reach the lower limit of the observational constraints \citep{Higuchi2017}.
If $T_\mathrm{chem}$ is lower than 100~K,
the high-density models yield the CO column density comparable to the observational constraints. 

The $Z$ dependence of ${\cal N}(\mathrm{CO})$
is shown in Figure \ref{fig:beta_comp}b.
For the fiducial models ($T_\mathrm{chem}=300$~K), 
 ${\cal N}(\mathrm{CO})$ decreases with increasing $Z$, 
roughly following 
the predictions from Equation (\ref{NCOpred_beta}).
When $T_\mathrm{chem}$ decreases to $10~$K, 
${\cal N}(\mathrm{CO})$
is increased by an order of magnitude, keeping 
the $Z$ dependence almost unchanged  (also see Figure \ref{fig:COfit}b).
Figure \ref{fig:beta_comp}b shows that metallicities larger than $Z=1$ cannot reproduce the observational CO column densities 
even when the H$_2$ formation is enhanced by decreasing $T_\mathrm{chem}$.

\subsubsection{49 Ceti}\label{sec:49Ceti}

    Following \citet{Hughes2017} and \citet{Higuchi2019},
    we adopt the disk model which has a power-law distribution with 
    the inner edge at $R_\mathrm{in}=20~\mathrm{au}$
    and with a exponential cut-off at $R=R_\mathrm{c}$, 
    \begin{equation}
        {\cal N}_\mathrm{C\perp}(R) \propto 
        \left( \frac{R}{R_\mathrm{c}} \right)^{-\gamma} 
        \exp\left[ - \left( \frac{R}{R_\mathrm{c}} \right)^{2-\gamma} \right],
        \label{49disk}
    \end{equation}
    where $R_\mathrm{in}=20~$au, $\gamma=-0.5$, 
    and $R_\mathrm{c}=140~$au are the 
    best fit parameters.
    The outer edge of the disk $R_\mathrm{out}$ is at infinity.
    The scale height is $h(R) = c_\mathrm{s}(R)/\Omega$, 
    where $c_\mathrm{s}$ is the sound speed with
    $T=23~\mathrm{K}(R/100~\mathrm{au})^{-1/2}$ and
    $\Omega = \sqrt{GM_\mathrm{star}/R^3}$,
    where $M_\mathrm{star}$ is  the central star mass
    \citep[][also see Table \ref{tab:49Ceti}]{Hughes2018}.
    
    As discussed in Section \ref{sec:modelparameter}, a typical mid-plane number density of carbon nuclei 
    is given by $\sim 10^3~\pcc$, where 
    ${\cal N}_\mathrm{C} \sim 10^{18}~\psc$ is divided by the disk extent 
    $R_\mathrm{c} - R_\mathrm{in}$.
    We consider two different values of the mid-plane carbon nucleus 
    density at $R=R_\mathrm{c}/2$, 
    $n_\mathrm{C,mid0}=4.9\times 
    10^2~\pcc$ (low-density model) 
    and $n_\mathrm{C,mid0}=2.0\times 10^3~\pcc$ (high-density model).

    \begin{table}
    \begin{center}
    \begin{tabular}{|l|c|c|}
        \hline
        & low-density & high-density \\
        \hline
        $n_\mathrm{C,mid0}$ [$\pcc$]$^{(1)}$ & $4.9\times 10^2$ & 
        $2.0\times 10^3$ \\
        \hline
        ${\cal N}_\mathrm{C}$ [$\psc$]$^{(2)}$ & $8.4\times 10^{17}$ & $3.3\times 10^{18}$ \\
        \hline
        ${\cal N}_\mathrm{C\perp}$ [$\psc$]$^{(3)}$ & $8.2 \times 10^{16}$ & $3.5\times 10^{17}$ \\
        \hline
        $f_\mathrm{dg}/f_\mathrm{dg,fid}$ & 0.24 & 0.06 \\
        \hline
    \end{tabular}
    \end{center}
    \caption{
    List of the model parameters for 49 Ceti.
    $^{(1)}$The mid-plane carbon nucleus number densities at $R=R_\mathrm{c}/2=60~$au. 
    $^{(2)}$The mid-plane carbon nucleus column densities integrated from the inner edge to infinity.
    $^{(3)}$The vertical carbon nucleus number densities with $Z=1$ at $R=60~$au.
    }
    \label{tab:49Ceti_num}
    \end{table}

   The mid-plane density is given by 
     \begin{eqnarray}
        n_\mathrm{C}(R,0) &=& 5.6\times 10^2 ~\mathrm{cm}^{-3} 
        \left( \frac{n_\mathrm{C,mid0}}{4.9\times 10^2~\pcc} \right) 
        \nonumber \\
        &  & \hspace{2mm} 
        \times \left( \frac{R}{60~\mathrm{au}} \right)^{-3/4} \exp\left[ - \left( \frac{R}{140~\mathrm{au}} \right)^{2.5} \right],
        \label{nCmid49}
    \end{eqnarray}
    where $\mu$ is assumed to be spatially constant.
    The corresponding mid-plane column densities are listed in Table \ref{tab:49Ceti_num}.
    The high-density model gives the mid-plane column density comparable to or 
    slightly larger than that predicted in \citet{Higuchi2019}.
    As in the case of $\beta$ Pictoris, the vertical column densities 
    are determined by assuming all the gas is neutral atomic, and 
    the values with $Z=1$ at $R=60~$au are 
    listed in Table \ref{tab:49Ceti_num}.

    Substituting ${\cal N}_\mathrm{C}$ into 
    Equation (\ref{fdg_determ}) gives 
    $\fracdg/\fracdgfid = 0.24$ for the low-density models and $0.06$ for the high-density models, where 
    $\taumid=10^{-2}$ is used. 
    Since $\fracdg/\fracdgfid$ is much lower than unity, 
    the acceleration of the CO formation owing to H$^*_2$
    is not effective (Section \ref{sec:CO1e17}).



    \begin{table}[htpb]
        \centering
        \begin{tabular}{|c|c|c|}
            \hline
            & SSUV & WSUV \\
            \hline
        $M_\mathrm{star}~[M_\odot]^{(1)}$ & $2.1$ & $2.0$ \\
            \hline
            $T_\mathrm{eff}~[\mathrm{K}]^{(2)}$ & $10,000$ & $9,000$ \\
            \hline
            $\log_{10} g^{(3)}$ & $4.5$ & $4.3$ \\
            \hline
        \end{tabular}
        \caption{
            $^{(1)}$ Stellar mass. $^{(2)}$ Effective temperature.  $^{(3)}$ Surface gravity.
            The stellar models of SSUV and WSUV 
            are taken from \citet{Hughes2008} and \citet{Roberge2013}, respectively.
            In both models, the metallicities [M/H] are assumed to be the solar one.
        }
        \label{tab:49Ceti}
    \end{table}

Different stellar parameters are used in the literature 
\citep{Chen2006,Hughes2008,Montesinos2009,Roberge2013}.
Among the different stellar models, 
the two extreme stellar models that provide the smallest and largest 
$\chi = \sqrt{\chi_\mathrm{co}\chi_\mathrm{oh}}$ are considered.
The stellar parameters of these models are summarized in Table \ref{tab:49Ceti}.
The strong and weak stellar UV models are called SSUV and WSUV, respectively.

\begin{figure*}[htpb]
        \centering
        \includegraphics[width=18.0cm]{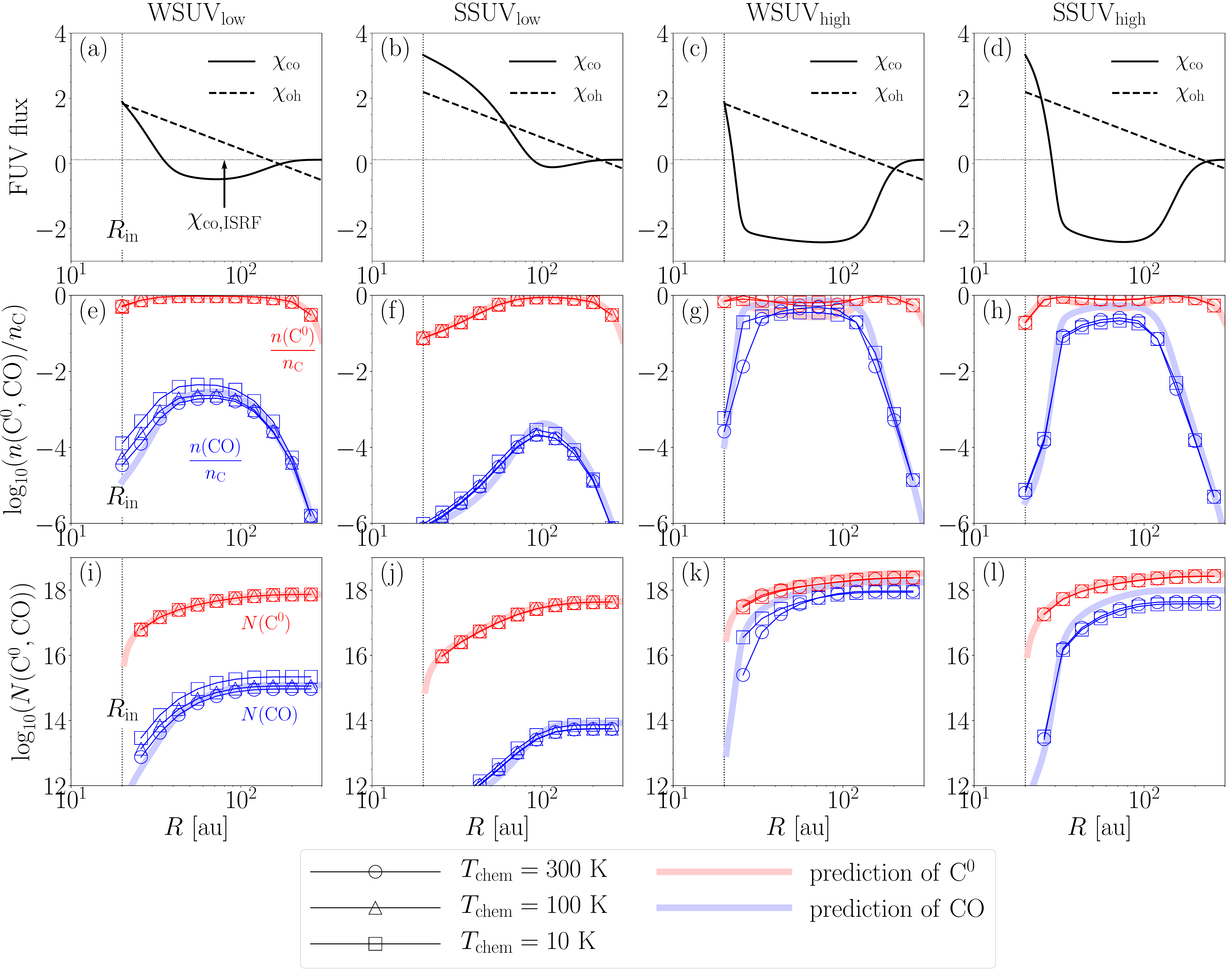}
        \caption{
             The same as Figure \ref{fig:betadisk} but for 49 Ceti.
             The vertical dashed lines show the position of the disk inner edge $R_\mathrm{in}=20$~au.
	}
\label{fig:49disk}
\end{figure*}

    Equating Equations (\ref{nCmid49}) and  Equation (\ref{nCreq}), 
    one obtains the expected CO fraction,
    \begin{eqnarray}
        \left(\frac{\nCO}{\nC}\right)_\mathrm{49Ceti} &=& 5.2 
        \times 10^{-4}
        \left( \frac{n_\mathrm{C,mid0}}{490~\pcc} \right)^{1.8}
        \left( \frac{R}{60~\mathrm{au}} \right)^{0.62} \nonumber \\
        & & \hspace{0mm} \times Z^{-1.1} \chico^{-1} 
        \left(\frac{\chioh}{\chi_\mathrm{oh,SSUV}}\right)^{-1} 
        \left( \frac{ {\cal A}_\mathrm{O}}{ 
        {\cal A}_\mathrm{O,ism}} \right) \nonumber \\
        && \hspace{0mm} \times \exp\left[ -1.8\left( \frac{R}{R_\mathrm{c}} \right)^{2.5} \right],
        \label{nCO49}
     \end{eqnarray}
     where $\chi_\mathrm{oh,SSUV}$ shows $\chioh$ of the SSUV model.

Models SSUV (WSUV) with the lower and higher mid-plane densities
are called 
SSUV$_\mathrm{low}$ (WSUV$_\mathrm{low}$) and 
SSUV$_\mathrm{high}$ (WSUV$_\mathrm{high}$), respectively.

First, the low-density models are focused on.
Figures \ref{fig:49disk}a and \ref{fig:49disk}b show the radial 
distributions of the normalized UV fluxes 
for models WSUV$_\mathrm{low}$ and SSUV$_\mathrm{low}$, respectively.
Since the C$^0$ fractions are smaller for SSUV owing to the 
stronger FUV flux, $\chico$ decreases 
more slowly for SSUV$_\mathrm{low}$ than for WSUV$_\mathrm{low}$.
As a result, more outer parts of the disk is exposed to 
intense FUV radiation for SSUV$_\mathrm{low}$.
The FUV fluxes $\chico$ take minimum values in $R\sim 60$~au 
for WUV$_\mathrm{low}$ 
and in $R\sim 10^2~$au for SUV$_\mathrm{low}$, 
and then increases to $\sim \chi_\mathrm{co,ISRF}$ 
around $R\sim 200~$au.
For WSUV$_\mathrm{low}$,
the vertical attenuation of $\chico$ is clearly seen 
in $40~\mathrm{au}<R<200~$au because 
${\cal N}_\mathrm{C,\perp}$ is larger than $N(\mathrm{C}^0)_\mathrm{shld}$ (Table \ref{tab:49Ceti_num}).



The radial profiles of the C$^0$ and CO fractions in 
the low-density models with are shown in 
Figures \ref{fig:49disk}e and \ref{fig:49disk}f.  
 For $T_\mathrm{chem}=300$~K,
both the models clearly show 
that CO forms efficiently only in the region where 
the stellar radiation flux is low because 
$(\nCO/\nC)$ is in proportion to $\chico^{-1}$.
Their radial distributions agree with the predictions from 
the analytic model (Section \ref{sec:COdiskana}).

A decrease in $T_\mathrm{chem}$ does not affect the CO 
fractions significantly, compared with the results with $\beta$ Pictoris
(Figures \ref{fig:49disk}e and \ref{fig:49disk}f).
The CO fractions are increased only by a factor of 2 or 3.
This comes from the fact that $\fracdg/\fracdgfid$ for 49 Ceti is more than one order of magnitude smaller than that for $\beta$ Pictoris.
Roughly speaking, 
the inner (outer) regions dominated by the stellar radiation (the vertical ISRF) correspond to the strong-FUV (weak-FUV) models
at $\NC \sim 10^{17}~\psc$.
As seen in the first and middle panels of Figure \ref{fig:COfit}, 
when $\fracdg < 0.24\fracdgfid$, the existence of H$_2^*$ increases the CO fractions only by a factor of 2 or 3.


When the mid-plane density increases,
the behaviors of the CO fractions change significantly.
The stellar radiation in the wavelength range of $\chico$ is
almost attenuated in very inner regions even for model SSUV$_\mathrm{high}$.
Since the vertical column density 
${\cal N}_\mathrm{C,\perp}=3.5\times 10^{17}~\psc$ at $R=60~$au (Table \ref{tab:49Ceti_num}) is 
much larger than $N(\mathrm{C^0})_\mathrm{shld}$,
the vertical attenuation of the ISRF is significant in 
$30~\mathrm{au}<R<200~\mathrm{au}$ (Figures \ref{fig:49disk}c and \ref{fig:49disk}d).

Figures \ref{fig:49disk}g  and \ref{fig:49disk}h show that 
both the high-density models yield large amounts of CO where $\chico$ is almost attenuated.
The CO fractions in the attenuated regions are almost independent of $T_\mathrm{chem}$.
This is consistent with the results shown in Section \ref{sec:CO2e18}. 
In a such attenuated region, 
$\mathrm{C^++H^*_2\rightarrow CH^+ + H}$ does not accelerate the CO formation because of depletion of C$^+$.
We should note that the CO fractions slightly decrease with decreasing $T_\mathrm{chem}$ 
in the attenuated regions
because the H atoms available for the CO chemistry shown in Figure \ref{fig:COnetwork} are reduced by enhancement of the H$_2$ formation (Section \ref{sec:CO2e18})


The bottom panels of Figures \ref{fig:49disk} show 
the radial distributions of $\NCI$ and $\NCO$ integrated from the inner edge to $R$.
Increases in the CO column densities are saturated around the peak radius of $\nCO$.
Thus, ${\cal N}_\mathrm{C}$ is determined in the amount of CO inside the peak radius of $\nCO$.

\begin{figure}[htpb]
\centering
\includegraphics[width=7cm]{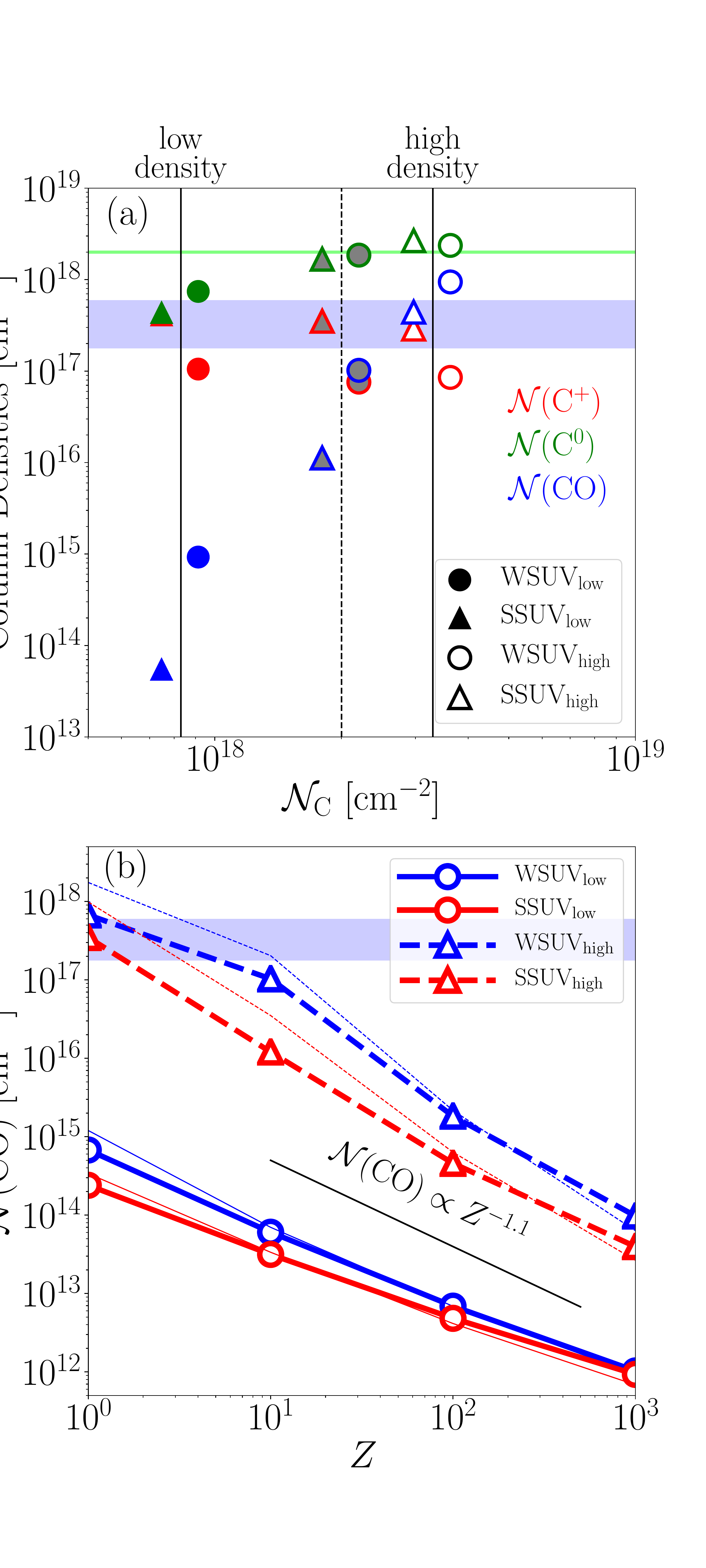}
\caption{
    Results of the PDR calculations for 49 Ceti.
(a) The mid-plane column densities of (red) $\mathrm{C}^+$, (green) $\mathrm{C}^0$, and (blue) CO 
as a function of ${\cal N}_\mathrm{C}$ for 
the models with the weak and strong stellar radiation models are shown by 
the circles and triangles, respectively.
The results with $T_\mathrm{chem}=300~$K are shown.
The results with the low and high density models are displayed by 
the filled and open markers.
For reference, the results with the intermediate density 
$n_\mathrm{C,mid0}=1.2\times 10^3~\pcc$ (${\cal N}_\mathrm{C}=2\times 10^{18}~\psc$)
are shown by the markers filled by gray.
At each ${\cal N}_\mathrm{C}$, 
the data points with different stellar models are slightly shifted horizontally for better visibility.
The green and blue rectangle regions indicate the observational constraints
on the C$^0$ and CO column densities obtained from \citet{Higuchi2019} and 
\citet{Higuchi2020}, respectively.
(b) The mid-plane CO column densities ${\cal N}(\mathrm{CO})$ as a function of $Z$ for (blue circles) WSUV$_\mathrm{low}$, 
(red circles) SSUV$_\mathrm{low}$, (blue triangles) WSUV$_\mathrm{high}$, and (red triangles) SSUV$_\mathrm{high}$.
For each model, the prediction from the analytic formula is shown by the thin line.
The blue rectangle region is the same as that in
Panel (a). As a reference, 
$\NCO\propto Z^{-1.1}$ is plotted
(Equation (\ref{nCO49})).
}
\label{fig:49_comp}
\end{figure}

The obtained C$^+$, C$^0$, and CO column densities integrated along the mid-plane are 
compared with the observational results in Figure \ref{fig:49_comp}a for $Z=1$.
The low-density models show that
the CO fractions does not reach the observed values 
for both the stellar models.
By contrast, the high-density models
produce sufficient amounts of CO to explain the observational results.
We thus expect that there is a best solution to explain the observed 
CO column densities in $4.9\times 10^2~\pcc<n_\mathrm{C,mid0}<2.0\times 10^3~\pcc$ 
(Table \ref{tab:49Ceti_num}).
For reference, we conducted additional PDR calculations with 
$n_\mathrm{C,mid0}=1.2\times 10^3~\pcc$ 
(${\cal N}_\mathrm{C}=2\times 10^{18}~\psc$).
The mid-plane density is between those in the low-density and high-density models.
Figure \ref{fig:49_comp}a show that a best solution is located 
in $1.2\times 10^3~\pcc<n_\mathrm{C,mid0}<2.0\times 10^3~\pcc$, and 
both ${\cal N}(\mathrm{C}^0)$
and ${\cal N}(\mathrm{CO})$ of a best solution 
will be consistent with the observational results.

The $Z$ dependence of ${\cal N}(\mathrm{CO})$ is shown in Figure \ref{fig:49_comp}b.
For all the models, the CO column densities decrease with increasing $Z$ as 
$ {\cal N}(\mathrm{CO}) \propto Z^{-1.1}$, and 
are consistent with the predictions from Equation (\ref{nCO49}). 
Figure \ref{fig:49_comp}b shows that
WSUV$_\mathrm{high}$ (SSUV$_\mathrm{high}$) provides ${\cal N}(\mathrm{CO})$ 
consistent with 
the observed CO column densities if $1< Z < 10$ ($Z\sim 1$).
This suggests that if the gas is secondary-origin ($Z\sim 10^{3}$), 
the CO column densities cannot be explained only by steady-state chemical reactions.

\subsection{ Uncertainty of the H$_2$ Formation on Dust Grains}\label{sec:ER}

One of the main results of this paper is that 
the CO formation is affected strongly by $T_\mathrm{chem}$, which is one of the uncertain 
parameters in the ER mechanism, when the shielding effects are not significant.
In this section, other uncertainties of the H$_2$ formation are discussed.

Another factor not considered in this paper is stochastic effects.  
Fluctuations of the number of H atoms on the grain surface 
should be take into account when 
few reactive H atoms are on the grain surface \citep{LePetit2009}.
We confirmed that the number density of the H atoms physisorbed on the grain surface is much 
smaller than the number density of dust grains.
This indicates that the master equation should be used instead of the rate equation 
to accurately estimate the H$_2$ formation rate of the LH mechanism.
\citet{Bron2014} found that the LH mechanism becomes an efficient H$_2$ formation mechanism 
in unshielded PDR regions even when the dust temperature is high for the interstellar environment.
It is still unknown how the fluctuations of the number of physisorbed H atoms affect H$_2$ formation 
in debris disks.

Fluctuations of the dust temperature also affects the H$_2$ formation 
for small grains with $a\le 0.02~\mu$m \citep{Cuppen2006}.
If only large dust grains exist in debris disks, dust temperature fluctuations are negligible 
because of their large heat capacities.
If small dust grains exist in debris disks, 
their temperature fluctuations may affect 
the H$_2$ formation rate.

\subsection{Caveats}\label{sec:caveat}

%
%
%

Our analytic formula of $\nCO$ enables us to predict the CO spatial distributions 
at a given $Z$ and stellar radiation field (Section \ref{sec:COdiskana}).
Comparison of our results with observational results gives a constraint on the gas metallicity. 
In order to confirm whether the steady-state chemistry is a promising mechanism to produce the 
observed amount of CO, it is crucial to constrain the gas metallicity by other methods.

We should note that the column density of hydrogen is 
constrained by several observations for the debris disk around $\beta$ Pictoris.
\citet{Freudling1995} obtained an upper limit of the 
HI column density, ${\cal N}(\mathrm{H^0})<
2-5\times 10^{19}~\mathrm{cm}^{-2}$ from non-detection of the 21~cm line.
\citet{Lecavelier2001} found that the H$_2$ column density should be smaller than
$\sim 10^{18}~\mathrm{cm}^{-2}$ because H$_2$ absorption lines are not detected.
Using a C$^0$ column density of $\sim (2.5\pm 0.7)\times 10^{16}~$cm$^{-2}$
\citep{Higuchi2017}, 
${\cal N}_\mathrm{C}/{\cal N}_\mathrm{H}$ is estimated to be larger than
$0.025$, which is much larger than ${\cal A}_\mathrm{C,ism}$.
Recently \citet{Matra2017} constrained the hydrogen number density by the line
ratio of CO ($J=2-1$) and CO ($J=3-2$); their result suggests that 
$\nH$ should be smaller than $\sim 10^4~\mathrm{cm}^{-3}$.
If this is the case, the observed CO fractions may not be explained by such a small value of 
$\nH$ even if $T_\mathrm{chem}$ is sufficiently low.
The combination of our results and the observational constraints
may rule out the primordial origin ($Z=1$) of the gas in $\beta$ Pictoris.

Although we found that there is a parameter set which may explain the amount of CO 
by considering only steady-state chemical reactions in Sections \ref{sec:betaPic} and \ref{sec:49Ceti}, 
we need to compare our models with observational results in more detail by 
performing synthetic observations of our disk models.

For 49 Ceti, \citet{Higuchi2020} showed that the excitation temperatures need to be as low as $\sim 10$~K
by conducting non-LTE analysis of the rotational spectral lines of CO considering 
isotopologues line ratios \citep[also see][]{Kospal2013}.
In our PDR calculations of 49 Ceti, the gas temperatures do not fall below $\sim 20~$K.
This low excitation temperature may suggest that level populations 
are not thermalized because of low $\nH$.

We will address direct comparisons between the chemical model and observational results 
in forthcoming papers.

\subsection{Dissipation of Protoplanetary Disks}

If the gas in debris disks contains a significant amount of H$_2$, the gas of PPDs
still remains, at least partially, in debris disks; i.e.
the timescale of PPD gas dispersal is longer than the age of debris disks.
Many authors investigated the gas dispersal of PPDs
through magnetically-driven winds \citep{Suzuki2009,Suzuki2014} or/and
photo-evaporation by UV/X-rays \citep{Hollenbach1994,GortiHollenbach2009,Owen2012,Nakatani2018}.
EUV photo-evaporation \citep{Hollenbach1994} 
is expected to be inefficient 
in debris disks around A-type 
stars because they emit less EUV than either lower 
and higher mass stars.
FUV photo-evaporation would be important at large radii $>100~$au 
as long as a large amount of small dust grains exist \citep{GortiHollenbach2009}.
During the evolution from a PPD to a debris disk, an amount of dust grains decreases and 
the grain size increases. 
\citet{Nakatani2021} recently found that 
this evolution of dust grains suppress 
photo-electric heating, making FUV photo-evaporation ineffective.
As a result, the lifetime of the gas component 
is determined by EUV photo-evaporation. 
Suppose that the EUV flux of an A type star is about $\sim 10^{39}~\mathrm{s}^{-1}$, the photo-evaporation is at least 
$3\times 10^{-11}~M_\odot~\mathrm{yr^{-1}}$ \citep{Gorti2009}.
If the initial disk mass at the point 
small grains have been depleted is 
$\sim 10^{-3}~M_\odot$, the lifetime is about $30~$Myr, 
indicating that the lifetime can be longer than $10~$Myr.
However, the EUV flux in young A-type stars is highly uncertain, and 
how protoplanetary disks evolve into small-grain-depleted disks are 
unknown. Further investigations are needed.
{The future high resolution observation for the ${\rm C}/{\rm CO}$ ratio
combined with Equation~(\ref{fitting}) would give the distribution of
$n_\mathrm{H}$ in depleting PPDs, which may reveal
the physics of disk dispersal.
}

\section{Summary}\label{sec:summary}
In this paper, we investigated the CO chemistry in optically thin PDRs with larger 
dust grains than the ISM as a model of debris disks.
Because hydrogen is involved in CO formation, the CO abundance should depend on the hydrogen 
density.
This may allow us to constrain the amount of hydrogen if
we know how much carbon nuclei becomes CO \citep{Higuchi2017}.

We investigated CO formation only through steady chemical reactions without
considering any secondary processes.
For simplicity, we consider a stationary plane-parallel semi-infinite uniform slab which
is illuminated by the interstellar standard radiation and
stellar radiation from its edge.
We compute the chemical and thermal equilibria taking into account detailed
radiative transfer using the Meudon PDR code \citep{LeP2006,Goi2007,LeBourlot2012}.
Although the plane-parallel geometry is far from the disk geometry, 
we found that the chemical structure of debris disks 
can be approximated by the findings from the one-dimensional plane-parallel PDR calculations.

The model parameters are the carbon nucleus density $\nC$,
gas metallicity $Z$, FUV photon number flux $\chico$ that dissociates CO normalized by the Habing flux,
the dust-to-gas mass ratio $\fracdg$,  and 
$T_\mathrm{chem}$, which corresponds to the energy barrier of chemisorption.
The model parameters are tabulated in Table \ref{tab:param}.
We call the models with weaker and strong FUV incident fluxes "weak-FUV" and "strong-FUV", respectively.
The dust-to-gas mass ratio is expressed in terms of the carbon nucleus column density at $\taumid=1$ ($\NCtaumid$), where
$\taumid$ is the mid-plane optical depth (Equation (\ref{fdg})).
Observations of debris disks can constrain $\NCtaumid$.
As a fiducial value of $\NCtaumid$, $\NCtaumidfid=2\times 10^{19}~\psc$ is adopted.
The corresponding dust cross section is 80 times smaller than that in the ISM. 
A fiducial dust-to-gas mass ratio $\fracdgfid$ is defined as $\fracdg$ at ${\cal N}_\mathrm{C}=\NCtaumidfid$.
The ratio $\fracdg/\fracdgfid$ is equal to  $(\NCtaumid/\NCtaumidfid)^{-1}$, and determines the dust surface area (Equation (\ref{crosssec})).
Considering large uncertainty in H$_2$ formation on the grain surface, 
we explore a range of H$_2$ formation rates on dust grains with
a dust temperature of $\sim 100~$K 
by adopting $T_\mathrm{chem}=300$~K and 10~K.
The $T_\mathrm{chem}=10~$K case gives an upper limit of the H$_2$ formation rate since there is almost no energy barrier in chemisorption.

Our results are summarized as follows:

\begin{enumerate}


          \item  For higher $T_\mathrm{chem}$ and  
           smaller $\fracdg$,  
           we found that CO formation proceeds in the H$_2$-poor
           environments without the influence of H$_2$. 
           Our environments tend to 
           have a very low H$_2$
    abundance because the low grain surface area, 
           high dust temperatures, and low gas temperatures make H2 
           formation extremely inefficient if the activation barrier 
           is as high as $T_\mathrm{chem}=300$~K.
           We developed an analytical formula for the CO fractions 
           (Section \ref{sec:analyticmodel}).
           The analytic formula is applicable even in 
           shielded regions by using the local flux attenuated 
           by C$^0$ and CO, and reproduces 
           the spatial distributions of the 
           CO fractions obtained from the PDR 
           calculations reasonably well (Figure \ref{fig:COshield}).
     

\item From the high density limit of
the analytic formula for $\nCO/\nC$ (Equation (\ref{fitting})), 
we obtain the hydrogen nucleus number
density required to produce a given $\nCO/\nC$ 
as a function of the FUV ﬂux and gas metallicity
(Equation (\ref{nHconstrained})). 
If the amount of carbon nuclei and 
local FUV flux are ﬁxed, 
lower metallicity produces a larger amount of CO
because there are more H atoms to start the 
chemistry that produces CO.

\item 
       The H$_2$ formation rate depends sensitively on $T_\mathrm{chem}$
       since it is proportional to $\exp(-T_\mathrm{chem}/T)$.
       The smaller $T_\mathrm{chem}$,
       the more active the H$_2$ formation on the grain surface.
       Increases in $\fracdg$ also enhances the H$_2$ formation.
          The CO formation is accelerated by vibrationally-excited H$_2$, whose internal energy is used to 
          overcome an endothermicity of $\mathrm{C^++H_2\rightarrow CH^+ + H}$.
          The enhancement of the CO fractions owing to the excited H$_2$ is occurred 
          only when the shielding 
          effects are insignificant.
          If the shielding effects are important, 
          the CO fractions show different dependence on 
          $T_\mathrm{chem}$ and $\fracdg$.
          As $T_\mathrm{chem}$ decreases  and/or 
          $\fracdg$ increases, the CO fractions decrease although the dependence is weak.
          This is because in such a shielded region, the C$^+$ fractions is too low for 
          $\mathrm{C^++H_2\rightarrow CH^+ + H}$ to accelerate the CO formation.
          A decrease in the number of H atoms available for CO formation decreases the CO fractions.
       
       \item 
           A critical dust-to-gas mass ratio 
           $(\fracdg/\fracdgfid)_\mathrm{cri}$ 
           above which the CO formation is accelerated depends 
           sensitively on $T_\mathrm{chem}$. 
           For $T_\mathrm{chem}=300~$K, roughly speaking,
           $(\fracdg/\fracdgfid)_\mathrm{cri}$ is about $\fracdgfid$
           although it depends on $\nC$ and $Z$ 
           (Section \ref{sec:parametersurvey}).
           A decrease of $T_\mathrm{chem}$ from 300~K to 10~K 
           reduces $(\fracdg/\fracdgfid)_\mathrm{cri}$ significantly;
           $(\fracdg/\fracdgfid)_\mathrm{cri} \lesssim 0.1$ for weak-FUV and $\sim 0.01$ for strong-FUV.
       

       \item 
       CO is self-shielded but also shielded by C$^0$.
Which shielding effect of CO 
           is effective depends on $\nCO/\nC$ at the low $\tau$ limit (Section \ref{sec:shielding}).
           For $\nCO/\nC > (\nCO/\nC)_\mathrm{cr} = 3\times 10^{-3}$ at $\tau\sim 0$, CO is self-shielded for
           $\NCO > 10^{14}~\psc$.
           When the CO fraction is sufficiently smaller than $(\nCO/\nC)_\mathrm{cri}$, 
           the C$^0$ attenuation increases the CO fraction for 
           $N(\mathrm{C}^0)>4\times 10^{16}~\psc$ before CO self-shielding becomes important.

       \item
       On the basis of the analytic formula shown in Section \ref{sec:analyticmodel}, 
       we developed a method to determine the spatial distributions of the CO number density in a given disk 
       structure by taking into account shielding effects 
       in the radiation field consistently in Section \ref{sec:COdiskana}.

\end{enumerate}

The chemical structure of debris disks have not been studied by 
taking into account detailed chemical processes consistent with 
radiation transfer. Particularly, it may be important to consider 
level populations of H$_2$ accurately 
because for 
lower $T_\mathrm{chem}$ and/or higher $\fracdg$ 
excited H$_2$ can promote endothermic 
formation reactions of CH$^+$ to accelerate the  CO formation. 

In this paper, an interstellar C/O ratio of 0.4 is used (Table \ref{tab:abundance}).
Recently \citet{Kama2016} and \citet{Bergin2016} however 
found evidence that C/O ratios could exceed unity
in PPDs, indicating that C/O ratios
change from the interstellar value during the evolution of PPDs.
If a part of the gas in a debris disk comes from a remnant PPD,
the C/O ratio could exceed unity also in a debris disk.
From our analytic formula for the CO fraction (Equation (\ref{fitting})),
it is expected that the CO fraction changes in proportion to ${\cal A}_\mathrm{O}$.

Our results were compared with the observational results of the 
debris disks around $\beta$ Pictoris and 49 Ceti 
in Sections \ref{sec:betaPic} and \ref{sec:49Ceti}, respectively.

For $\beta$ Pictoris as an example of gas-depleted debris disks, 
the steady chemical model is difficult to produce a sufficient amount of CO to fit the observational results 
if the fiducial parameter $T_\mathrm{chem}=300~$K 
of the Eley-Rideal mechanism 
for the formation of H$_2$ on warm dust grains
is adopted. 
The higher the metallicity, the more inefficient 
CO formation becomes, due to the lack of H atoms to 
facilitate its formation.
The CO column densities decrease as $\propto Z^{-1}$.
If $T_\mathrm{chem}$ is lower than 100~K, 
the steady chemical model may explain the observed amount of CO 
due to the H$_2$-accelerated CO formation 
only if $Z\sim 1$ (Figure \ref{fig:beta_comp}).
If $Z>1$, the steady chemical model cannot produce a sufficient amount of CO even when the H$_2$ formation on dust surfaces are efficient.

For 49 Ceti as an example of gas-rich debris disks, 
the CO fraction is sensitive to the column density of carbon nuclei (Figure \ref{fig:49_comp}).
Our results show that there is a solution that may explain 
both the observed CO and C$^0$ column densities \citep{Higuchi2020} 
in a reasonable range of ${\cal N}_\mathrm{C}$ 
for $Z=1$ \citep{Higuchi2019}.
Since the dust-to-gas mass ratio 
is extremely small and 
shielding effects are significant, 
the CO fraction is insensitive to $T_\mathrm{chem}$.
As in the case of $\beta$ Pictoris, the observed CO column density 
cannot be reproduced for 49 Ceti if $Z>10$.

We should note that 
steady state chemical models can explain $\beta$ Pictoris and 49~Ceti only
under a very restricted set of parameters,
including enhanced production of H$_2$ on grains and $Z=1$. 
However, for $\beta$ Pictoris, 
complementary published observations mentioned in Section \ref{sec:caveat}
indicated much lower H or H$_2$ densities than are required. 
The combination of this work and the
constraints on the H and H$_2$ reveals that steady state chemistry cannot explain the
observations.

As mentioned in Section \ref{sec:intro},
if the gas in $\beta$ Pictoris is secondary-origin and 
CO is supplied from solid bodies through collisions,
a required injection rate of CO is estimated by the observed CO mass divided by 
the CO photo-dissociation timescale.
Our results clearly show that the CO formation rate with $Z\gtrsim 1$ 
is too small to explain the observed amount of CO.
The total solid bodies required to provide CO during the stellar age 
is about $30~M_\otimes$ taking into  account an expected fraction of CO in the solid bodies
\citep[also see][]{Dent2014}.
As pointed out in Section \ref{sec:intro}, 
we need to consider detailed outgassing processes from solid bodies 
in order to investigate if the required amount of CO can be supplied.
In addition, the gas should be removed at a timescale comparable to the CO photo-dissociation 
timescale because C$^0$ and C$^+$ are accumulated if the gas is not removed. 
Viscous disk evolution \citep{Kral2017}, 
EUV photo-evaporation \citep{Nakatani2021}, and disk winds may be important to remove the gas.



\begin{acknowledgments}
We thank the reviewers for providing us many constructive comments that improve this paper signficantly.
We thank Gianni Cataldi, Satoshi Yamamoto, Nami Sakai, Munetake Momose, Kenji Furuya, and Masanobu Kunitomo for valuable discussions.
This work was supported in part by JSPS KAKENHI Grant Numbers 21H00056 (K.I.),
22H00179, 22H01278, 21K03642, 18H05436, 18H05438 (H.K.),
18K03713, 19H05090 (A.E.H.), 18H05222, 20H05844, 20H05847 (Y.A.).
Y.A. also acknowledges support by NAOJ ALMA Scientific Research Grant code 2019-13B.

\end{acknowledgments}

{\software{ Meudon PDR code \citep{LeP2006}, numpy \citep{Numpy},
Matplotlib \citep{Matplotlib}}}

%





\appendix

\section{ An Analytic Formula for the CO Fraction in 
the Simplified Chemical Network without Hydrogen Molecule}\label{app:COana}

In this appendix, we construct an analytical model for CO formation by the simplified 
CO formation network without H$_2$ shown in Figure \ref{fig:COnetwork}.
With the rate coefficients listed in Table \ref{tab:reaction},
the four chemical balance equations for CO, CO$^+$, OH, and CH$^+$ are solved, where 
$n(e^-)$ is equal to $\nCp$, 
$n(\mathrm{H})$ is set to be $\nH$ in the H$_2$-free environment, most oxygen is in 
the atomic form ($n(\mathrm{O}) \sim n_\mathrm{O}$), and the C$^0$ and C$^+$ fractions are obtained 
from Equation (\ref{nCIana}) with $\nCp = \nC - n(\mathrm{C}^0)$, where 
$\nCO\ll \nCI , \nCp$ is assumed.
To investigate the contribution of each path to the CO fraction, the analytical form of the CO fraction 
is divided into three parts,
\begin{equation}
        \nCOana \equiv \nCO_\mathrm{oh} +\nCO_\mathrm{chp} +  \nCO_\mathrm{woH},
  \label{COana}
\end{equation}
where $\nCO_\mathrm{oh}$, $\nCO_\mathrm{chp}$, and $\nCO_\mathrm{woH}$ are the CO fractions formed 
through the three pathways.

\begin{table}[htpb]
	\centering
	\begin{tabular}{|l|l|}
    \hline
		chemical reaction & rate coefficient \\
		\hline
		\hline
                $\mathrm{C^+ + H\rightarrow CH^+ + h\nu}$ & 
                $k_\mathrm{cp,h}=2.29\times 10^{-17}(T/300)^{-0.42}$ \\
                $\mathrm{CH^+ + H\rightarrow H_2 + C^+}$ & 
                $k_\mathrm{chp,h}=7.5\times 10^{-10}$ \\
                $\mathrm{CH^+ + O\rightarrow CO^+ + H}$ & 
                $k_\mathrm{chp,o}=3.5\times 10^{-10}$ \\
    $\mathrm{O + H\rightarrow OH + h\nu}$ & $k_\mathrm{o,h}=9.9\times 10^{-19} (T/300)^{-0.38}$ \\
    $\mathrm{OH + C^+ \rightarrow H + CO^+}$ & $k_\mathrm{oh,cp}=7.7\times 10^{-10} (T/300)^{-0.5}$ \\
    $\mathrm{OH + C \rightarrow H + CO}$ & $k_\mathrm{oh,c}=1.0\times 10^{-10}$ \\
    $\mathrm{CO^+ + e^- \rightarrow C + O}$ & $k_\mathrm{cop,e}=4.58\times 10^{-7}$ \\
    $\mathrm{CO^+ + H \rightarrow CO + H^+}$ & $k_\mathrm{cop,h}=7.5\times 10^{-10}$ \\
    $\mathrm{OH + h\nu \rightarrow O + H}$ & $\alpha_\mathrm{oh}=4.3\times 10^{-6}\chioh$ \\
    $\mathrm{CO^+ + h\nu \rightarrow C^++O}$ & $\alpha_\mathrm{cop}=1.35\times 10^{-8}\chico^{1.3}$ \\
    $\mathrm{CO + h\nu \rightarrow C+O}$ & $\alpha_\mathrm{co}= 10^{-10}\chico$ \\
    \hline
	\end{tabular}
	\caption{
    Chemical reaction rates that are important in the CO formation shown in Figure \ref{fig:COnetwork}.
    The rate coefficients in the first eight reactions are shown 
    in an unit of cm$^3$~s$^{-1}$.
    The last three rate coefficients correspond to the unattenuated photo-dissociation rates in s$^{-1}$.
	}
	\label{tab:reaction}
\end{table}

\begin{figure}[htpb]
  \centering
  \includegraphics[width=15cm]{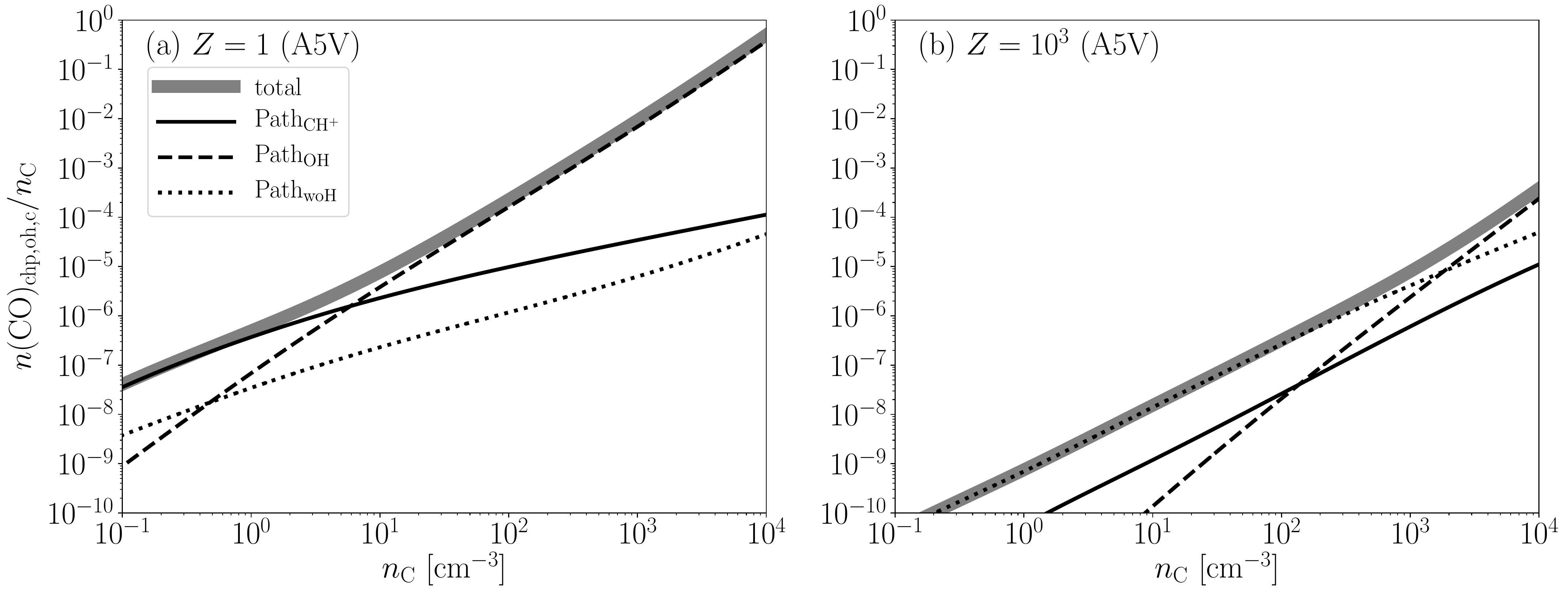}
  \caption{
      The contributions of the three different paths to the analytic CO fraction as a function 
      of $\nC$ for (a)$Z=1$ and (b)$Z=10^3$.
  }
  \label{fig:ana_z}
\end{figure}

    The contributions of the three CO formation paths to the CO fraction for $Z=1$ and $Z=10^3$ 
    are illustrated in Figures \ref{fig:ana_z}a and \ref{fig:ana_z}b, respectively. 
    The spectral type is fixed to A5V.
    Which CO formation path is important depends both on the gas metallicity and the gas density.
    For $Z=1$, CO forms mainly from CH$^+$ for low densities (Figure \ref{fig:ana_z}a).  
    Path$_\mathrm{OH}$ becomes the dominant pathway to form CO 
    when $\nC$ exceeds $\sim 10~\pcc$. 

    As $Z$ increases, 
    both $\nCO_\mathrm{chp}$ and $\nCO_\mathrm{oh}$ decrease at a fixed $\nC$ 
    since $\nH$ decreases with $Z$.
    By contrast, $\nCO_\mathrm{woH}$ is independent of $Z$ at a fixed $\nC$ because hydrogen 
    is not involved in \Pathwoh.
    As shown in Figure \ref{fig:ana_z}b, 
    for $Z=10^3$, $\nCO_\mathrm{woH}$ dominates over $\nCO_\mathrm{chp}$ by an order of magnitude, and 
    {\Pathwoh} becomes the dominant pathway to form CO for lower densities.
    When $\nC>10^3~\pcc$, the CO formation through {\Pathoh} determines the CO fraction.


%

We focus on the high density limit where {\Pathoh} is dominant.
The path $\mathrm{O}\rightarrow \mathrm{OH}\rightarrow \mathrm{CO}$ 
contributes to the CO formation more than 
$\mathrm{O}\rightarrow \mathrm{OH}\rightarrow\mathrm{CO^+}\rightarrow \mathrm{CO}$.
Using the fact that the main destruction processes of OH and CO are photo-dissociation, 
we obtain the CO fraction as follows:
The OH abundance is determined by the balance between the radiative association 
$\mathrm{O+H\rightarrow OH+h\nu}$ 
(the rate coefficient $\koh$) and the OH photo-dissociation (the rate coefficient $\aOH$).
Eventually, OH is combined with C to form CO with a rate coefficient of $\kohc$.
Considering the CO photo-dissociation (the rate coefficient $\aCO$), 
the analytical form of the CO fraction is given by 
\begin{eqnarray}
    \frac{n(\mathrm{CO})_\mathrm{oh}}{\nC} &=& \frac{\koh\kohc}{\aCO\aOH}\nO\nH \nonumber \\
   &=& 7.4\times 10^{-17}~T_{300}^{-0.38}~\frac{\Ao}{ {\cal A}_\mathrm{O,ism}} \zeta^2,
\label{COana0}
\end{eqnarray}
where
\begin{equation}
    \zeta = \frac{\sqrt{Z}\nH}{\chi},\;\;\;\mathrm{where}\;\;\chi = \sqrt{\chico\chioh},
\end{equation}
$T_{300}=T/300~\mathrm{K}$, 
${\cal A}_\mathrm{O}$ is the oxygen abundance of the gas phase
and ${\cal A}_\mathrm{O,ism} = 3.2\times 10^{-4}Z$ 
(Table \ref{tab:abundance}),  
and we use the fact that $\aCO\propto \chico$ and $\aOH \propto \chioh$.
Equation (\ref{COana0}) shows that 
the CO fraction depends on $\nH Z^{0.5}\chi^{-1}$, which is quite similar to $\eta$.
The difference between $\eta$ and $Z^{0.5}\nH \chi^{-1}$ comes from the 
temperature dependence shown in Equation (\ref{COana0}).

\begin{figure}[htpb]
  \centering
  \includegraphics[width=7cm]{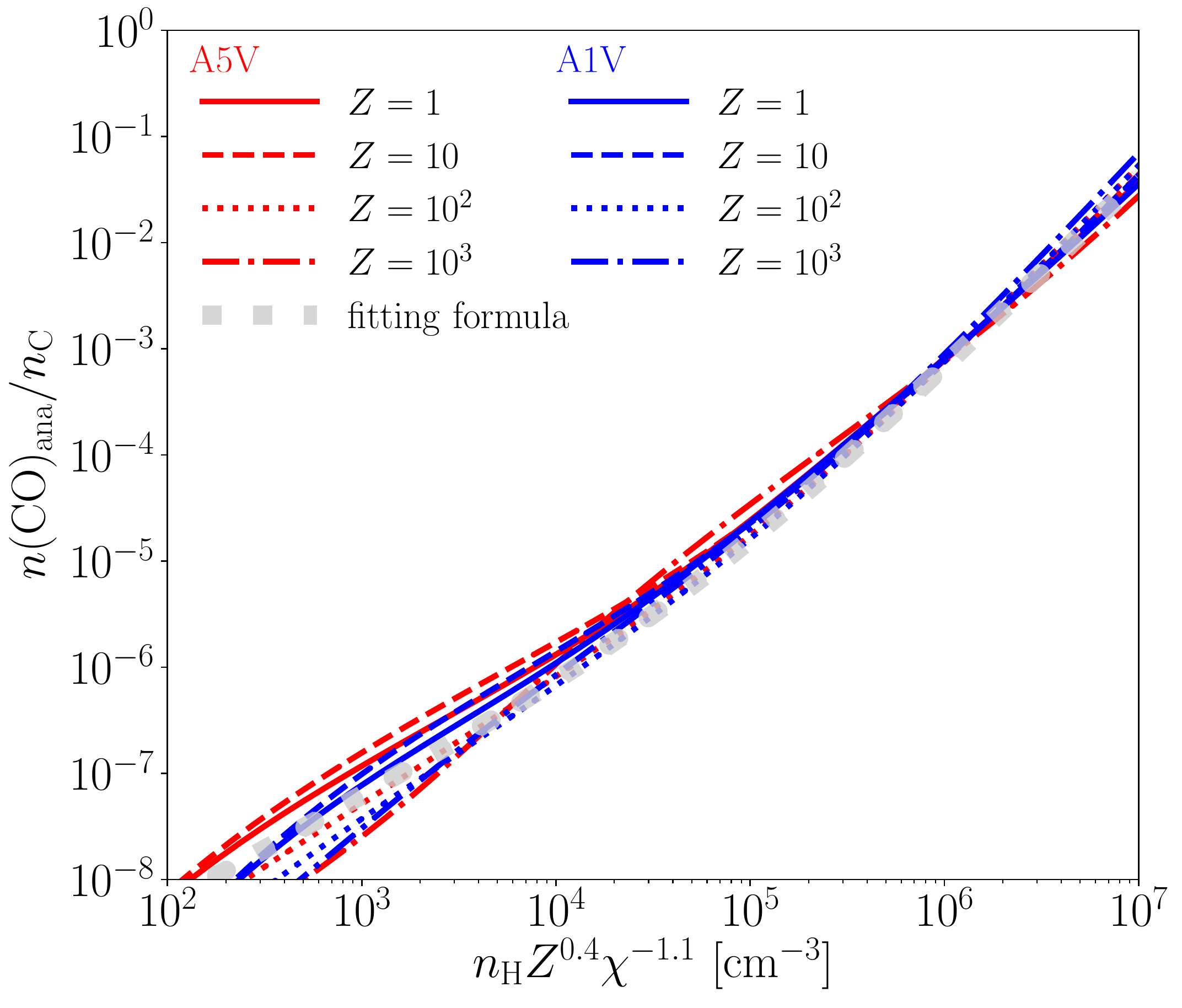}
  \caption{
      The CO fraction estimated by Equation (\ref{COana}) for various metallicities and spectral types. 
      The horizontal axis is $\eta = \nH Z^{0.4}\chi^{-1.1} = \zeta (Z\chi)^{-0.1}$.
      The gray dotted line corresponds to the fitting function shown by Equation (\ref{fitting0}).
  }
  \label{fig:COana}
\end{figure}

Equation (\ref{COana0}) shows that $\nCOana$ is expected to be characterized by $\zeta$ at least for high densities
where $\nCO_\mathrm{ana}\sim \nCO_\mathrm{oh}$. 
In order to check this, we fit $\nCOana$ with a function of $\eta = \nH Z^a \chi^{b}$, 
where $a$ and $b$ are the fitting parameters.
As shown in Figure \ref{fig:COana}, we found that 
$\nCOana$ is insensitive to both $Z$ and $\chi$  for $\eta >10^{5}~\pcc$ 
if $\eta$ with $a=0.4$ and $b=-1.1$ is taken as the horizontal axis.
The difference between $\zeta$ and $\eta$ with $a=0.4$ and $b=-1.1$ comes from 
the negative temperature dependence of $\nCO_\mathrm{oh}$ (Equation (\ref{COana0})).
The gas temperatures increase when either $Z$ or $\chi$ increaes.
Even for low densities where {\Pathoh} is no longer dominant CO formation path,
$\nCOana$ is characterized by $\eta=\nH Z^{0.4}\chi^{-1.1}$ reasonably well. 
A fitting function of the CO fraction is given by 
\begin{equation}
    \frac{\nCO_\mathrm{ana}}{\nC} = \frac{\Ao}{ {\cal A}_\mathrm{O,ism}} 
    \left(
        10^{-14} \eta^{1.8}
    + 6.0\times 10^{-11} \eta \right).
    \label{fitting0}
\end{equation}

\end{document}